\documentclass[journal]{IEEEtran}
\usepackage[latin1]{inputenc}
\usepackage{graphicx}
\usepackage{color}
\usepackage{placeins}
\usepackage{float}
\usepackage{tabularx,colortbl}
\usepackage{amssymb}
\usepackage{amsthm}
\usepackage{cite}
\usepackage{amsmath}
\usepackage{caption2}
\usepackage{cleveref}
\usepackage{multicol}       
\usepackage{multirow}       
\usepackage{array}          
\usepackage{makecell}
\usepackage{colortbl}
\usepackage{pifont}       
\usepackage{bbding}       
\usepackage{fontawesome}  
\usepackage{cases,subeqnarray}
\usepackage{bm,multirow,bigstrut}
\usepackage{textcomp}
\usepackage{latexsym,bm}
\usepackage{booktabs}
\usepackage{xcolor}
\usepackage{mathtools}
\usepackage{dsfont}
\usepackage{extarrows}

\usepackage{subfigure}
\usepackage{makecell}
\definecolor{lightblue}{rgb}{0.93,0.95,1.0}

\usepackage{algorithm} 
\usepackage{algorithmic}

\theoremstyle{plain}

\theoremstyle{plain}

\theoremstyle{definition}

\newcommand{\ra}[1]{\renewcommand{\arraystretch}{#1}}

\IEEEoverridecommandlockouts

\begin{document}
\title{Deep Unfolding Beamforming and Power Control Designs for Multi-Port Matching Networks}
\author{Bokai~Xu, Jiayi~Zhang,~\IEEEmembership{Senior Member,~IEEE}, Qingfeng~Lin, Huahua Xiao,\\Yik-Chung~Wu,~\IEEEmembership{Senior Member,~IEEE}, and Bo~Ai,~\IEEEmembership{Fellow,~IEEE}
%
%
\thanks{B. Xu, J. Zhang, and B. Ai are with the School of Electronic and Information Engineering, Beijing Jiaotong University, Beijing 100044, China.}
\thanks{H. Xiao is with ZTE Corporation, State Key Laboratory of Mobile Network and Mobile Multimedia Technology.}
\thanks{Q. Lin and Y. Wu are with the Department of Electrical and Electronic Engineering, The University of Hong Kong, Hong Kong.}
}
\maketitle

\vspace{-1.75cm}
\begin{abstract}
The key technologies of sixth generation (6G), such as ultra-massive multiple-input multiple-output (MIMO), enable intricate interactions between antennas and wireless propagation environments. As a result, it becomes necessary to develop joint models that encompass both antennas and wireless propagation channels. To achieve this, we utilize the multi-port communication theory, which considers impedance matching among the source, transmission medium, and load to facilitate efficient power transfer.
Specifically, we first investigate the impact of insertion loss, mutual coupling, and other factors on the performance of multi-port matching networks.
Next, to further improve system performance, we explore two important deep unfolding designs for the multi-port matching networks: beamforming and power control, respectively.
For the hybrid beamforming, we develop a deep unfolding framework, i.e., projected gradient descent (PGD)-Net based on unfolding projected gradient descent.
For the power control, we design a deep unfolding network, graph neural network (GNN) aided alternating optimization (AO)-Net, which considers the interaction between different ports in optimizing power allocation.
Numerical results verify the necessity of considering insertion loss in the dynamic metasurface antenna (DMA) performance analysis. Besides, the proposed PGD-Net based hybrid beamforming approaches approximate the conventional model-based algorithm with very low complexity.
Moreover, our proposed power control scheme has a fast run time compared to the traditional weighted minimum mean squared error (WMMSE) method.
\end{abstract}

\IEEEpeerreviewmaketitle
\vspace{-0.3cm}
\section{Introduction}
The rapid advancement of wireless communication systems, e.g., beyond fifth-generation (B5G) and sixth-generation (6G), has intensified the need for higher data rates, reliability, global coverage, and ultradense connectivity \cite{9390169,xu2023resource}.
Among the technologies is massive multiple-input multiple-output (MIMO), which has become a key enabling technology in 5G and future generations of wireless networks \cite{wang2023tutorial}.
Motivated by the potential benefits it offers, novel research directions are emerging under various names, such as holographic MIMO \cite{9136592}, large intelligent surfaces (LIS) \cite{10556753} and dynamic metasurface antenna (DMA) \cite{[46], [47]}.

Further advancement of MIMO technology will require not only accounting for radiative near-field effects, but also comprehensively modeling antenna interactions within an array. Consequently, communication theory and electromagnetic (EM) theory need to be closely integrated. By establishing a multi-port circuit theory model for wireless communication systems, communication circuit theory aims to bridge the gap between electromagnetic physics and information theory mathematics \cite{1310320}.

Since multi-port matching networks are capable of better assessing near-field channel characteristics, it is important to design efficient near-field signal processing algorithms for them, including optimization of DMA beamforming design, hybrid beamforming, and corresponding power allocations.
The challenge of optimizing a phase-shifter-based hybrid beamforming lies in the non-convex constant modulus constraints, the strong coupling relationship between analog beamforming and digital beamforming matrices, and the power limitations in multi-port matching networks. Consequently, efficient hybrid beamforming methods that address these challenges have garnered significant attention in the literature, encompassing conventional model-based optimizations as well as pure data-driven deep learning (DL) approaches \cite{10475378, liu2023cellfree}.

While such purely data-driven methods are promising in many applications, they have the following drawbacks: long training times, huge memory requirements, and virtually no performance guarantees \cite{8715338, 10697206}. Therefore, the use of hybrid model-based DL methods is becoming more common in communications systems instead of purely data-driven methods. Among these hybrid model-based DL methods, deep unfolding \cite{9020494} is a powerful example that is rapidly gaining traction in the communications industry.

\subsection{Prior Works}
A first attempt at modelling mutual coupling between antenna circuits and the propagation environment can be found in \cite{1167256}. Moreover, the paper \cite{1310320} developed an electrical network model of the MIMO communication system considering a single user.
Despite its great potential, the investigation of beamforming capabilities and transmission rates of multi-port matching networks is still in infancy, primarily because of the absence of realistic models.
Although these effects are relevant, they are only partially considered, and the basic beamforming techniques in \cite{PhysRevApplied.8.054048} are derived without considering backward propagation in the waveguides, insertion losses, and mutual coupling.

On the other hand, \cite{9127817} employed multi-port communication theory to investigate the reciprocity between uplinks and downlinks, as well as the mutual information among users.
A recent study \cite{10096991} examined the impact of mutual coupling on single-user MIMO channel estimation.
Furthermore, \cite{damico2024holographic} studied the effects of antenna spacing and mutual coupling on channel gain and spectral efficiency (SE) using multi-port theory. It was found that as antenna distance decreases, SE increases because more antennas allow more energy to be collected rather than because mutual coupling occurs.

Another candidate antenna architecture is based on metasurface antennas for mMIMO, whose implementation in signal processing is still an area of active research. 
Similarly, the authors in \cite{[58], 10001283} developed a multi-port channel model for DMA that included the propagation and reflections of waves throughout waveguides that supply the antenna elements, as well as the mutual coupling of waves and waveguides.
Additionally, in multi-port matching networks, conducting a comprehensive performance analysis and signal processing design is vital for a deep understanding of inter-user interference (IUI) and for further enhancing SE, which, to the best of our knowledge, has not yet been adequately reported in prior works. 

Additionally, we focus on the hybrid beamforming and power control issues of multi-port impedance matching networks. With a limited number of radio frequency (RF) chains, hybrid beamforming can offer a cost- and energy-effective multiplexing solution \cite{10286447}. 
Specifically, in order to maximize the SE of the multi-port matching networks, both the digital precoder and analog precoder should be optimized jointly, a highly non-convex problem, which would result in significant computational overhead and a higher likelihood of infeasible solutions.
Traditional optimization schemes, such as manifold optimization based alternating minimization (MO-AltMin) algorithm \cite{7397861}, can effectively address this problem. However, they need faster convergence speed and lower complexity.

Recently, DL has gained much attention in wireless communications for its ability to deal with complex and challenging problems. This is primarily attributed to its ability to handle complex and challenging problems by extracting valuable features within high-dimensional spaces.
Two typical DL techniques are often applied: purely data-driven DL and hybrid model-based DL \cite{shlezinger2023aiempowered,9893021}.
The former uses deep neural networks (DNNs), convolutional neural networks (CNNs), or deep reinforcement learning to generate hybrid beamformers.
However, data-driven black box neural networks necessitate a substantial volume of training samples, incurring significant training costs and posing challenges in terms of interpretability and generalizability, thereby limiting their ability to ensure consistent performance.
An efficient inference mapping can be achieved by combining domain knowledge with data using model-based DL.

The deep-unfolding neural network is designed using the projected gradient and alternating direction method of multipliers (ADMM) algorithms.
Deep unfolding treats problem-specific algorithms' iteration as machine learning model layers. We can perform finite iterations on traditional optimization algorithms, introduce new parameters, and train them on the data using the established gradient descent strategy \cite{10286447}. Specifically, through deep unfolding with learned hyperparameters, the iterative optimizer's functionality is entirely retained, ensuring both its adaptability and comprehensibility. Compared with traditional neural network architectures, this approach of combining expert knowledge with machine learning concepts can give unfolding-based architectures better generalization performance and interpretability \cite{9020494, shlezinger2023aiempowered}.
Therefore, the resulting unfolded algorithm with learned parameters can effectively address various challenges within communication systems.

Essentially, the power control is the problem of optimizing a specific system-level utility function, e.g., sum-rate SE, maximum-min SE. Although this paradigm has shown remarkable success, many optimization problems it formulates are nonconvex and NP-hard, posing significant challenges in finding feasible solutions.
Recent studies \cite{9403959, 9944643} have demonstrated that DL-based methods outperform traditional approaches in wireless resource allocation. Generally, these tasks can be divided into two categories, i.e., supervised learning and unsupervised learning.
As an example, a supervised deep network has been used to approximate the classical weighted minimum mean squared error (WMMSE) model \cite{5756489} for power allocation in interference channels \cite{8444648}. 

While supervised learning continues to be a powerful and widely used method for reaching wireless resource allocation, its performance is limited by the variability of the training data. Furthermore, acquiring and preparing labeled data can be time-consuming, costly, and sometimes impossible, especially when dealing with large datasets. Model-based DL utilizes a supervised or unsupervised neural network and directly incorporates the optimization objective as a loss function.
In \cite{9403959}, the author used a deep unfolding network to expand the traditional WMMSE algorithm and designed an efficient model driven power allocation algorithm.
However, most of the research scenarios described above concern millimeter wave technologies, so beamforming and power control should be designed considering the multi-port matching network.

\subsection{Our Contributions}
To fill in the above gap, in this paper, we investigate the effects of insertion loss and mutual coupling on multi-port network systems and develop deep unfolding algorithms to design beamforming and power control. 
We try to answer two important questions: \emph{what phenomena inherent to multi-port matching networks should be considered to render realistic results?} and \emph{how can efficient signal processing algorithms be implemented to improve the system performance of multi-port matching networks, including beamforming and power control?}
Our contributions are summarized as follows. 
\begin{itemize}
\item Using the multi-port communication theory, we evaluate the performance of MIMO systems using the linear signal processing scheme. In the DMA system, traditional models commonly employed in MIMO become invalid, necessitating the consideration of electromagnetic phenomena such as mutual coupling, insertion losses, and reflections occurring within the waveguides. Unlike mMIMO systems, DMAs have a maximum SE, and increasing the number of antennas in each waveguide does not improve performance after reaching a certain threshold. Furthermore, DMA has a higher energy efficiency (EE) than traditional antenna structures.

\item We propose an unfolding framework, i.e., projected gradient descent (PGD)-Net, to design hybrid architectures based on the multi-port network system.
In contrast to most existing DL-aided beamforming designs, the unfolding framework delves into the matrix approximation problem associated with hybrid beamforming design rather than focusing on maximizing SE. 
The network architecture based on deep unfolding provides rapid convergence even with limited training data and without requiring training labels.

\item We then focus on power control issues in multi-port matching network systems. We propose an unfolded network of WMMSE, alternating optimization (AO)-Net, for power allocation in the multi-port network system, where the network weights are learned utilizing the graph neural network (GNN). The proposed method is unsupervised in that it does not require the solution of any power allocation problems for training. In spite of this, our approach still incorporates part of classical methods into the learning architecture of the deep unfolding network. Our objective is to utilize expert knowledge in developing theoretical models that achieve nearly optimal performance while significantly reducing execution time.
    
\item Simulation results demonstrate that the PGD-Net scheme outperforms conventional MO-AltMin approaches in terms of time efficiency and computational complexity.
Furthermore, our proposed GNN-aided AO-Net power control scheme offers a possibility for optimizing the performance-complexity trade-off.

\end{itemize}

\subsection{Organization and Notations}
The rest of this paper is organized as follows. 
The system under consideration and the circuital model are introduced in Section \uppercase\expandafter{\romannumeral2}.
Sections \uppercase\expandafter{\romannumeral3} and \uppercase\expandafter{\romannumeral4} detail the proposed PGD-Net and AO-Net designs for beamforming and power allocation, respectively.
Section \uppercase\expandafter{\romannumeral5} presents the numerical results. 
Finally, Section \uppercase\expandafter{\romannumeral6} concludes the paper by summarizing the main findings and discussing future directions.

\emph{{Notations}}: $\mathbb{C}^{M\times M}$ denotes the set of $M$-by-$M$ complex numbers. The boldface lower-case letters $\bf{x}$ and boldface upper-case letters $\bf{X}$ represent the column vectors and matrices, respectively. The subscripts $\mathbf{X}^{*}$, $\mathbf{X}^{T}$, and $\mathbf{X}^{H}$ denote the conjugate, the transpose, and the conjugate transpose of the matrix $\mathbf{X}$, respectively. $\operatorname{Re}\{\cdot\}$, $\operatorname{Tr}\{\cdot\}$ and $\mathbb{E}\left[\cdot\right]$ denote the real part, the trace operator and the expectation operator, respectively.
$\left \| {\bf X} \right \|_{F}  =\sqrt{\left \langle {\bf X},{\bf X} \right \rangle } $ represents the Frobenius norm of the matrix $\bf X$ and ${\bf I}_{N}$ is an $N\times N$ identity matrix.
$\circ $ is the Hadamard product.
$n\sim \mathcal{C N}\left(0, \sigma^{2}\right)$ is a complex Gaussian distribution with mean $0$ and covariance $\sigma$.
\begin{figure}[t]
	\centering
	\includegraphics[width=0.45\textwidth]{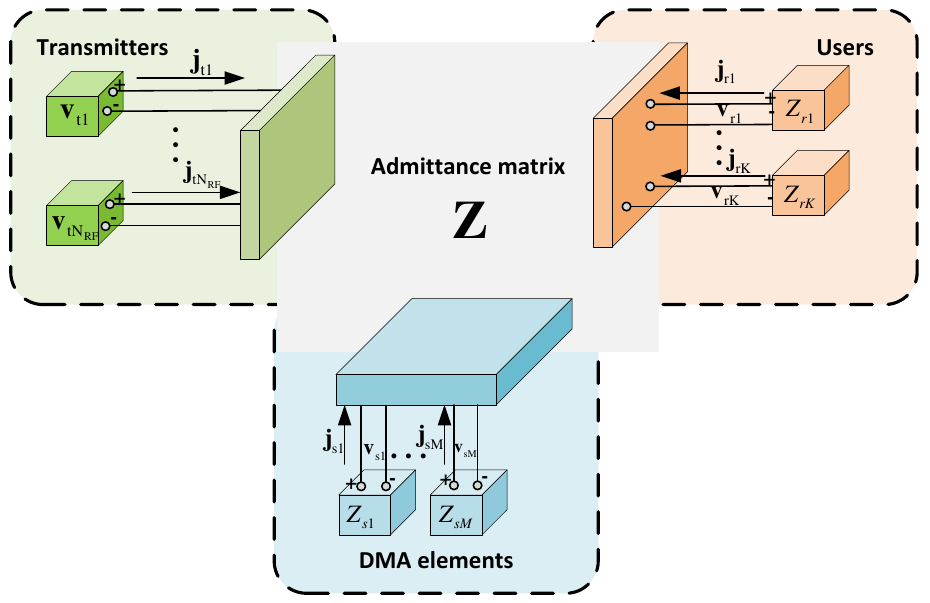}
	\caption{Circuit multi-port model for DMA based system. The ports situated on the left denote the system's input, correlating with the $N_{\text{RF}}$ distinct RF chains supplying the DMA. The ports on the right represent the UEs which are terminated in load impedances $Z_{rk}$.} \label{Fig.NearReg}
    \label{Compare_scenarios}
\end{figure}

\section{System Model}
We consider a downlink multi-user MIMO system in which one base station (BS) is deployed on a LIS and serves $K$ user equipments (UEs) within a single cell. The users are all regarded as magnetic dipoles, while the BS is equipped with full-digital (FD) arrays, hybrid digital/analog arrays or DMAs.
The DMA consists of $N_{\text{RF}}$ one-dimensional waveguides, each having a length of $L_{\mu }$, with $N_{\mu }$ small antenna elements embedded within each waveguide. Therefore, the number of antennas is represented by $M=N_{\text{RF}}\times N_{\mu}$. Each waveguide is connected to a dedicated RF chain. Additionally, each antenna element is equipped with a semiconductor device, enabling intentional tuning and admittance alteration.

The multi-port communication theory employs a circuit-theoretic methodology, where the inputs and outputs of a multi-antenna communication system correspond to ports of a multi-port black box characterized by impedance matrices. The entire MIMO transceiver system can be represented as a noisy multi-port circuit featuring the mutual MIMO impedance matrix.
As illustrated in Fig. 1, we can construct the above system into a multi-port matching network. Considering that the interaction solely depends on the admittance matrix $\mathbf{Z}$ between the transmitter and currents in different ports, it allows for including diverse effects within this matrix. In addition to mutual coupling, DMA elements and waveguides might interact, scattering between receivers and transmitters may occur, and there may be different channel conditions for wireless communication. Therefore, the whole network is characterized by
\begin{equation}
\left[\begin{array}{c}
\mathbf{v}_{\text{t}} \\
\mathbf{v}_{\text{s}} \\
\mathbf{v}_{\text{r}}
\end{array}\right]=\underbrace{\left[\begin{array}{lll}
\mathbf{Z}_{\text{tt}} & \mathbf{Z}_{\text{st}}^{T} & \mathbf{Z}_{\text{rt}}^{T} \\
\mathbf{Z}_{\text{st}} & \mathbf{Z}_{\text{ss}} & \mathbf{Z}_{\text{rs}}^{T} \\
\mathbf{Z}_{\text{rt}} & \mathbf{Z}_{\text{rs}} & \mathbf{Z}_{\text{rr}}
\end{array}\right]}_{\mathbf{Z}}\left[\begin{array}{l}
\mathbf{j}_{\text{t}} \\
\mathbf{j}_{\text{s}} \\
\mathbf{j}_{\text{r}}
\end{array}\right].
\end{equation}
The magnetic currents entering the transmitter are represented by $\mathbf{j}_{\text{t}} \in \mathbb{C}^{N_{\text{RF}} \times 1}$, while $\mathbf{j}_{\text{s}} \in \mathbb{C}^{M\times 1}$ and $\mathbf{j}_{\text{r}} \in \mathbb{C}^{K \times 1}$ denote the DMA radiating elements and the users' currents, respectively.
Similarly, $\mathbf{v}_{\text{t}} \in \mathbb{C}^{N_{\text{RF}} \times 1}$, $\mathbf{v}_{\text{s}} \in \mathbb{C}^{M \times 1}$ and $\mathbf{v}_{\text{r}} \in \mathbb{C}^{K \times 1}$ represent the vectors encompassing the magnetic voltages across their respective ports.

And the different admittance matrices $\mathbf{Z}_{\text{t}\text{t}^{\prime}}$ capture the coupling between the corresponding ports.
Specifically, $\mathbf{Z}_{\text{tt}} \in \mathbb{C}^{N_{\text{RF}} \times N_{\text{RF}}}$ denotes the coupling between different transmitter ports, $\mathbf{Z}_{\text{ss}} \in \mathbb{C}^{M \times M}$ captures the interaction among the DMA elements, and $\mathbf{Z}_{\text{rr}} \in \mathbb{C}^{K \times K}$ indicates the interaction among users. On the other hand, $\mathbf{Z}_{\text{st}} \in \mathbb{C}^{M \times N_{\text{RF}}}$ denotes the coupling that exists between each transmitter and the reconfigurable components, while $\mathbf{Z}_{\text{rt}} \in \mathbb{C}^{K \times N_{\text{RF}}}$ denotes the coupling between RF chains and users. Finally, $\mathbf{Z}_{\text{rs}} \in \mathbb{C}^{K \times M}$ denotes the wireless propagation environment between each radiating element and users.

Based on the application of Ohm's law, the following equations are obtained
\begin{equation}
\mathbf{v}_{\text{s}}=-\mathbf{Z}_{\text{s}} \mathbf{j}_{\text{s}},
\end{equation}
\begin{equation}
\mathbf{v}_{\text{r}}=-\mathbf{Z}_{\text{r}} \mathbf{j}_{\text{r}}.
\end{equation}
The negative sign is attributed to the direction of current flowing from the negative terminal to the positive terminal.
By combining equations (2) and (3) with the first two rows of equation (1), making use of the assumption $\mathbf{Z}_{\text{rt}}= \mathbf{0}$, and isolating the receiver currents, we obtain the following expressions
\begin{equation}
\begin{aligned}
\mathbf{j}_{\mathrm{r}}= & \left(\mathbf{Z}_{\text{r}}+\mathbf{Z}_{\text{rr}}-\mathbf{Z}_{\text{rs}}\left(\mathbf{Z}_{\text{s}}+\mathbf{Z}_{\text{ss}}\right)^{-1} \mathbf{Z}_{\text{rs}}^{T}\right)^{-1} \\
& \times\left(\mathbf{Z}_{\text{rs}}\left(\mathbf{Z}_{\text{s}}+\mathbf{Z}_{\text{ss}}\right)^{-1} \mathbf{Z}_{\text{st}}-\mathbf{Z}_{\text{rt}}\right) \mathbf{j}_{\text{t}}.
\end{aligned}
\end{equation}
In addition to the DMA characteristics, (4) is dependent on the wireless channel $\mathbf{Z}_{\text{rs}}$ and the terminating admittances $\mathbf{Z}_{\text{s}}$ and $\mathbf{Z}_{\text{r}}$.
In addition to incorporating the wireless channel, $\mathbf{Z}_{\text{rs}}$ encompasses the path loss, so that this term diminishes as twice the path loss or, in other words, as the distance between users and the BS.
The currents received by the users outside the reactive near-field can be simplified as
\begin{equation}
\mathbf{j}_{\text{r}}=\left(\mathbf{Z}_{\text{r}}+\mathbf{Z}_{\text{rr}}\right)^{-1}\left(\mathbf{Z}_{\text{rs}}\left(\mathbf{Z}_{\text{s}}+\mathbf{Z}_{\text{ss}}\right)^{-1} \mathbf{Z}_{\text{st}}-\mathbf{Z}_{\text{rt}}\right) \mathbf{j}_{\text{t}},
\end{equation}
where $\mathbf{Z}_{\text{s}} \in \mathbb{C}^{M \times M}$ and $\mathbf{Z}_{\text{r}} \in \mathbb{C}^{K \times K}$ are diagonal matrices with elements $\left(\mathbf{Z}_{\text{s}}\right)_{l, l}=Z_{s l}$ and $\left(\mathbf{Z}_{\text{r}}\right)_{m, m}=Z_{r m}$.
We assume that LIS employs linear superposition to combine multiple information-carrying patterns $\left\{\mathbf{f}_{k}\right\}_{k=1}^{K}$ for coherent transmission, thus the combined current distribution  $\mathbf{j}_{\text{t}}$ on the LIS can be modeled as $\mathbf{j}_{\text{t}}=\sum_{k=1}^{K}\mathbf{f}_{k}s_{k}$, where the beamforming matrix $\mathbf{F}=\left[\mathbf{f}_{1}, \mathbf{f}_{2}, \ldots, \mathbf{f}_{K}\right] \in \mathbb{C}^{M \times K}$ and the vector of symbols intended for $K$ users $\mathbf{s}=\left[s_{1}, s_{2}, \ldots, s_{K}\right]^{T} \in \mathbb{C}^{K \times 1}$ are involved. It is assumed that $\mathbb{E}\left[\mathbf{s} \mathbf{s}^{H}\right]=\sigma_{x}^{2} \mathbf{I}_{K}$, where $\sigma_{x}^{2}$ denotes the variance and $\mathbf{I}_{K}$ represents the identity matrix of size $K$. 
Then, the received signal is denoted by
\begin{equation}
\mathbf{y}=\mathbf{H} \mathbf{Fs}+\mathbf{n}
\end{equation}
with the noise term $\mathbf{n} \sim \mathcal{C N}\left(\mathbf{0}, \sigma_{n}^{2} \mathbf{I}_{K}\right)$.
Based on the relationship between the current at the transmitter and the current at the UE in (5), the equivalent channel can be denoted as
\begin{equation}
\mathbf{H}=\tilde{\mathbf{Z}}_{\text{r}}\left(\mathbf{Z}_{\text{rs}}\left(\mathbf{Z}_{\text{s}}+\mathbf{Z}_{\text{ss}}\right)^{-1} \mathbf{Z}_{\text{st}}-\mathbf{Z}_{\text{rt}}\right)
\end{equation}
with $\tilde{\mathbf{Z}}_{\text{r}}=\sqrt{\frac{\operatorname{Re}\left\{\mathbf{Z}_{\text{r}}\right\}}{2}}\left(\mathbf{Z}_{\text{r}}+\mathbf{Z}_{\text{rr}}\right)^{-1}$.
Furthermore, the received signal of the $k$-th UE is
\begin{equation}
y_{k}=\mathbf{h}_{k}^{H} \sum_{i=1}^{K} \mathbf{f}_{i} s_{i}+n_{k}, k=1,2, \ldots, K
\end{equation} 
with the equivalent channel $\mathbf{H}=\left[\mathbf{h}_{1}, \mathbf{h}_{2}, \ldots, \mathbf{h}_{K}\right]^{H}$ and the noise $n_{k}\sim \mathcal{C N}\left(0, \sigma_{n}^{2}\right)$.
The signal-to-interference-and-noise ratio (SINR) of the $k$-th UE is given by
\begin{equation}
\gamma_{k}=\frac{\left|\mathbf{h}^{H}_{k}\mathbf{f}_{k}\right|^{2}}{\sum_{m=1, m \neq k}^{K}\left|\mathbf{h}^{H}_{k}\mathbf{f}_{m}\right|^{2}+\frac{\sigma_{n}^{2}}{\sigma_{x}^{2}}}.
\end{equation}
Consequently, the SE of the system is
\begin{equation}
{\sf SE}=\sum_{k=1}^{K} \log _{2}\left(1+\gamma_{k}\right).
\end{equation}
In this case, we can compute the transmitted power as \cite{pozar2011microwave}
\begin{equation}
P_{t}=\frac{1}{2} \mathbb{E}\left[\operatorname{Re}\left\{\mathbf{j}_{\text{t}}^{H} \mathbf{v}_{\text{t}}\right\}\right]=\frac{\sigma_{x}^{2}}{2} \operatorname{Tr}\left\{\operatorname{Re}\left\{\mathbf{F}^{H} \mathbf{Z}_{\text{p}} \mathbf{F}\right\}\right\},
\end{equation}
where $\mathbf{Z}_{\text{p}}$ is the admittance matrix at the transmitter. We define $\mathbf{v}_{\text{t}}=\mathbf{Z}_{\text{p}} \mathbf{j}_{\text{t}}$. According to (1), (2) and (3), $\mathbf{Z}_{\text{p}}$ can be written as
\begin{equation}
\mathbf{Z}_{\text{p}}=\mathbf{Z}_{\text{tt}}-\mathbf{Z}_{\text{st}}^{T}\left(\mathbf{Z}_{\text{s}}+\mathbf{Z}_{\text{ss}}\right)^{-1} \mathbf{Z}_{\text{st}}.
\end{equation}

\begin{figure}[t]
	\centering
	\includegraphics[width=0.35\textwidth]{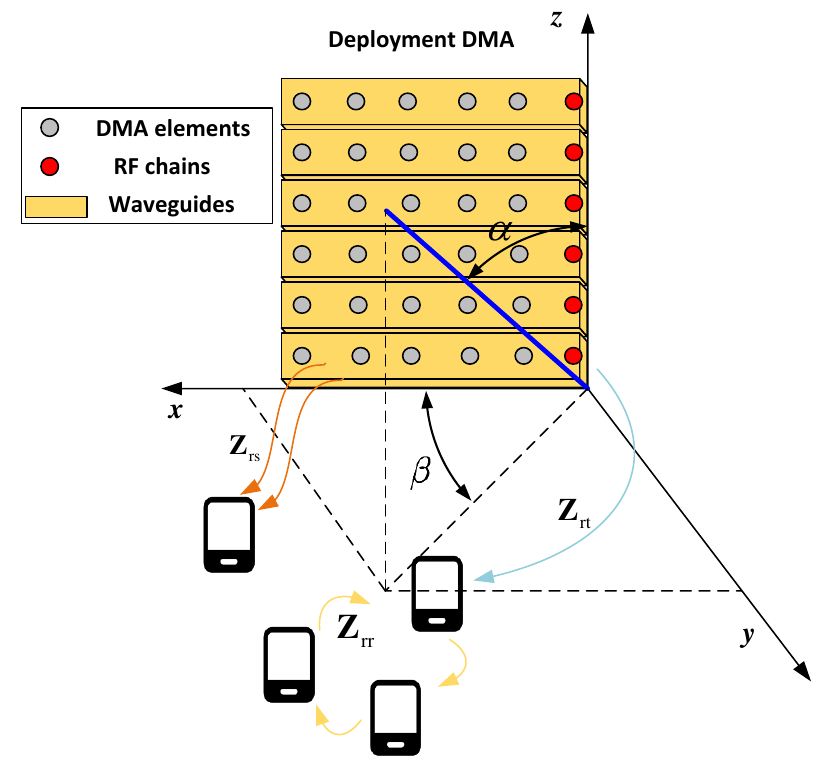}
	\caption{The system model and the DMA structure are depicted in graphical form. The polar angle domain is defined as $\alpha \in \mathbb{R}: 0 \leq \alpha \leq \pi$, while the azimuth angle domain is defined as $\beta \in \mathbb{R}: 0 \leq \beta < 2\pi$.
} \label{DMA}
\end{figure}

On the other hand, since the matrix $\mathbf{Z}_{\text{rs}}$ accounts for the wireless channel, it encapsulates the impact of the propogation medium. Hence, we write the admittance matrix $\mathbf{Z}_{\text{rs}}$ as
\begin{equation}
\mathbf{Z}_{\text{rs}}=\left[\begin{array}{llll}
\mathbf{z}_{1} & \mathbf{z}_{2} & \cdots & \mathbf{z}_{K}
\end{array}\right]^{T},
\end{equation}
where $\mathbf{z}_{k} \in \mathbb{C}^{M \times 1}$ is the channel vector of the $k$-th user. Under far-field conditions, i.e., when each radiating element in the DMA has the same angle of departure, the mutual admittance to the $k$-th user can be expressed as \cite{9300189}
\begin{equation}
\mathbf{z}_{k}=\sum_{l=1}^{L_{p}} \frac{\varphi _{l}}{\sqrt{L_{p}}} \sin \left(\alpha _{l}\right) \sin \left(\zeta _{l}\right) \mathbf{a}_{t}\left(\alpha _{l}, \beta _{l}\right),
\end{equation}
where $L_{p}$ is the number of paths. $\alpha_{l}$ and $\zeta_{l}$ denote the polar angles of departure and arrival, respectively.
The random variable $\varphi _{l}$ is an independent and identically distributed complex variable. It models the pathloss, phase, and polarization-shift of the $l$-th path, with a zero mean. 
Accordingly, the polar and azimuth angles are defined as in Fig. \ref{DMA}.
Moreover, the steering vectors $\mathbf{a}_{t}\left(\alpha _{l}, \beta _{l}\right) \in \mathbb{C}^{M \times 1}$ are given by
\begin{equation}
\mathbf{a}_{t}\left(\alpha _{l}, \beta _{l}\right)=\left[\begin{array}{lll}
\mathrm{e}^{i \mathbf{k}\left(\alpha _{l}, \beta _{l}\right) \mathbf{r}_{1}} & \cdots & \mathrm{e}^{i \mathbf{k}\left(\alpha _{l}, \beta _{l}\right) \mathbf{r}_{M}}
\end{array}\right]^{T}
\end{equation}
with $\mathbf{r}_{m}$ for $m=1, \ldots, M$ is the vector position of the $m$-th DMA element and
\begin{equation}
\mathbf{k}(\alpha, \beta )=k\left[\begin{array}{lll}
\sin (\alpha) \cos ( \beta) & \sin (\alpha) \sin (\beta) & \cos (\alpha)
\end{array}\right].
\end{equation}
As the number of samples $L_{p}$ tends to infinity, the random vector $\mathbf{z}_{k}$ follows a circularly symmetric complex Gaussian distribution denoted as $\mathcal{C \mathcal { N }}\left(\mathbf{0}, \boldsymbol{\Sigma}_{k}\right)$. 
And the covariance matrix $\boldsymbol{\Sigma}_{k}$ can be denoted by
\begin{equation}
\boldsymbol{\Sigma}_{k}=\varepsilon_{\varphi}^{2} \mathbb{E}_{\alpha, \beta}\left[\sin (\alpha)^{2} \mathbf{a}_{t}(\alpha, \beta ) \mathbf{a}_{t}(\alpha, \beta )^{H}\right] \mathbb{E}_{\zeta}\left[\sin (\zeta )^{2}\right],
\end{equation}
where $ \varepsilon_{\varphi}^{2}$ denotes the variance of the random variables $\varphi_{l}$, which is expressed as
\begin{equation} 
\varepsilon_{\varphi}^{2}=\left(\frac{2 \omega \epsilon}{4 \pi \left\|\mathbf{r}_{m}-\mathbf{r}_{k}\right\|_{2}}\right)^{2} S_{p},
\end{equation}
where $\mathbf{r}_{m}$ for $m=1, \ldots, M$ and $\mathbf{r}_{k}$ for $k=1, \ldots, K$ denote the position vectors corresponding to the DMA elements and the UEs, respectively.
And $ \omega $ is the angular frequency, $\epsilon$ is the permittivity of the medium, $S_{p}$ is the power losses due to polarization mismatch.
In the following, we provide detailed generic circuit model for DMA, FD mMIMO, and hybrid mMIMO systems, respectively.


\subsection{DMA System Model}
When the antenna at the BS is the DMA structure, we consider a model wherein each dedicated RF chain connects to multiple stacked one-dimensional waveguides. The transmitter ports in (1) no longer correspond to the antennas at the BS, but rather denote the output of the RF chains that supply the waveguides forming the structure.
By considering $\mathbf{Z}_{\text{rt}}=\mathbf{0}$ in (5), the equivalent channel can be expressed as
\begin{equation}
\mathbf{H}_{\text{dma}}=\tilde{\mathbf{Z}}_{\text{r}}\left(\mathbf{Z}_{\text{rs}}\left(\mathbf{Z}_{\text{s}}+\mathbf{Z}_{\text{ss}}\right)^{-1} \mathbf{Z}_{\text{st}}\right),
\end{equation}
where $\mathbf{Z}_{\text{s}} \in \mathbb{C}^{M \times M}$ is a diagonal matrix comprising the adjustable load admittances associated with each element of the DMA.
The tuning process entails modifying the imaginary component of $\mathbf{Z}_{\mathrm{s}}$, whereas the real component $\operatorname{Re}\left\{\left(\mathbf{Z}_{\text{s}}\right)_{i, i}\right\}=R_{s}\space \space\forall i$ signifies the parasitic resistance.
In order to fully characterize the system, it is not sufficient to consider only the transmitted power. The supplied power, denoted as $P_s$, is also required, which is given by the parameter
\begin{equation}
P_{s}=\frac{\sigma_{x}^{2}}{2} \operatorname{Tr}\left\{\operatorname{Re}\left\{\mathbf{F}^{H}\left(\mathbf{I}_{M}-\boldsymbol{\Lambda }^{H} \boldsymbol{\Lambda }\right)^{-1} \mathbf{Z}_{\text{p}} \mathbf{F}\right\}\right\},
\end{equation}
where
\begin{equation}
\boldsymbol{\Lambda }=\left(\mathbf{Z}_{\text {in }}-\mathbf{I}_{M} Z_{\text{0}}\right)\left(\mathbf{Z}_{\text {in }}+\mathbf{I}_{M} Z_{\text{0}}\right)^{-1}\in \mathbb{C}^{M \times M}
\end{equation}
is a diagonal matrix comprising the reflection coefficients.

We introduce $Z_{0}$ as the designated term for the source's characteristic impedance and the input admittance in the waveguides is $\mathbf{Z}_{\text {in }}=\mathbf{Z}_{\text{p}} \circ \mathbf{I}_{N_{\text{RF}}}$. Then (20) can be simplified as
\begin{equation}
P_{s}=\frac{\sigma_{x}^{2}}{2} \operatorname{Tr}\left\{\operatorname{Re}\left\{\mathbf{F}^{H} \mathbf{Z}_{\text{q}} \mathbf{F}\right\}\right\},
\end{equation}
where
\begin{equation}
\mathbf{Z}_{\text{q}}=\left(\mathbf{I}_{N_{\text{RF}}}-\boldsymbol{\Lambda}^{H} \boldsymbol{\Lambda}\right)^{-1} \mathbf{Z}_{\text{p}}.
\end{equation}
\subsection{FD and Hybrid mMIMO}
In the FD case, the ports corresponding to the DMA elements vanish in (1), and each of the transmitters directly represents one antenna attached to a dedicated RF chain. The relation between the currents and voltages is therefore given by
\begin{equation}
\left[\begin{array}{l}
\mathbf{v}_{\text{t}} \\
\mathbf{v}_{\text{r}}
\end{array}\right]=\underbrace{\left[\begin{array}{ll}
\mathbf{Z}_{\text{tt}} & \mathbf{Z}_{\text{rt}}^{T} \\
\mathbf{Z}_{\text{rt}} & \mathbf{Z}_{\text{rr}}
\end{array}\right]}_{\mathbf{Z}}\left[\begin{array}{l}
\mathbf{j}_{\text{t}} \\
\mathbf{j}_{\text{r}}
\end{array}\right].
\end{equation}
Each antenna is modeled as a magnetic dipole on an ideal electric conductor (PEC). We set $M=0$, therefore, the equivalent channel is
\begin{equation}
\mathbf{H}_{\text{fd}}=-\widetilde{\mathbf{Z}}_{\text{r}} \mathbf{Z}_{\text{rt}},
\end{equation}
where $\mathbf{Z}_{\text{rr}} \in \mathbb{C}^{K \times K}$ represents the mutual coupling between users, and $\mathbf{Z}_{\text{rt}} \in \mathbb{C}^{K \times N_{\text{RF}}}$ denotes the wireless propagation channel.
Moreover, when neglecting backscattering, we observe in this scenario that $\mathbf{Z}_{\text{p}} \approx \mathbf{Z}_{\text{tt}}$ \cite{[58]}.

Moreover, we consider a fully-connected topology for the hybrid analog-to-digital (A/D) architecture.
Consequently, the only distinction lies in the division of the beamforming matrix $\mathbf{F}$ into the analog matrix $\mathbf{F}_{\text{A}}$ and the digital matrix $\mathbf{F}_{\text{D}}$.

\section{Beamforming Design in Multi-Port Matching Networks}
In this section, we propose two kinds of beamforming design: DMA beamforming and hybrid beamforming. For DMA beamforming, we apply a gradient descent approach that considers the optimization of insertion loss \cite{10001283} in DMA systems. For hybrid beamforming, we utilize deep unfolding networks based on the traditional PGD algorithm, which is a commonly used iterative optimization algorithm.
\begin{figure*}[t]
	\centering
	\includegraphics[width=0.85\textwidth]{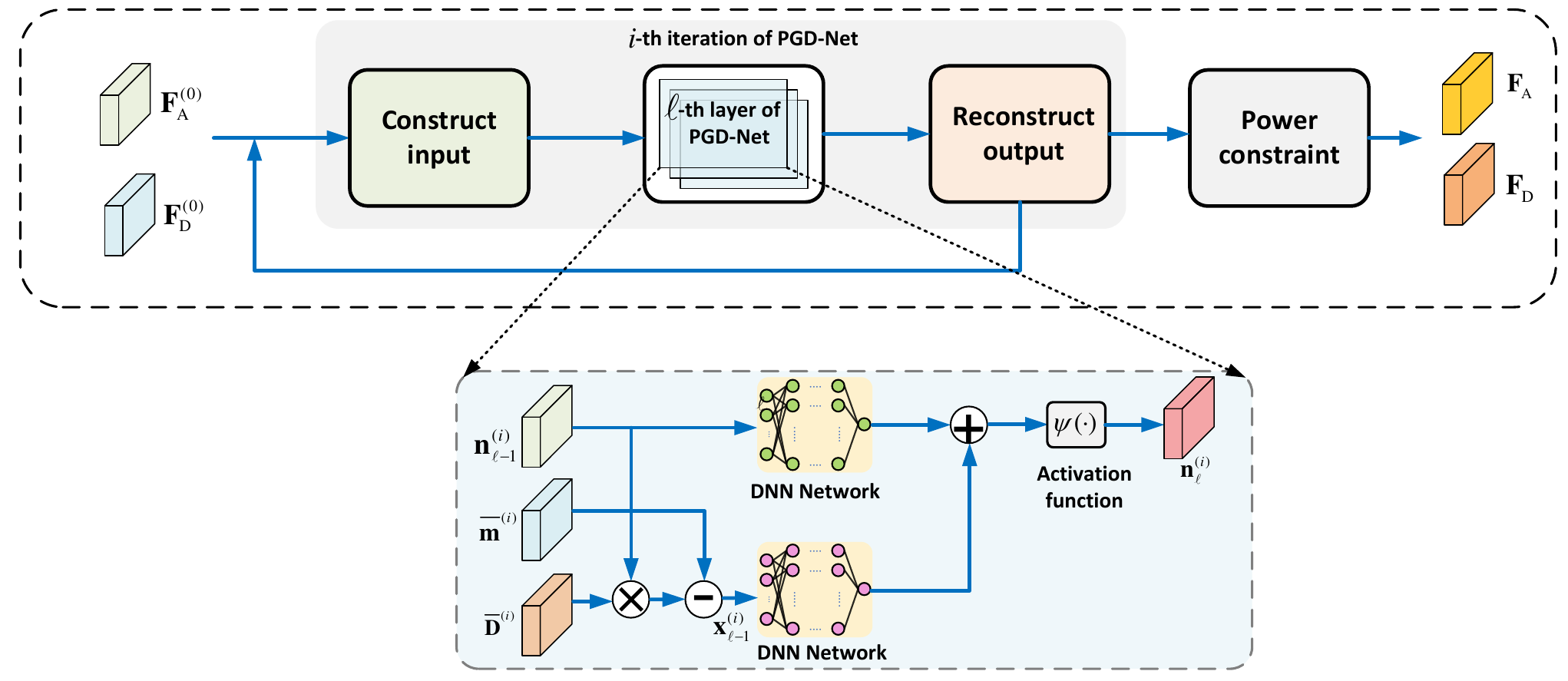}
	\caption{The architecture of the proposed PGD-Net for hybrid beamforming design. The blue box represents the specific architecture of the $\ell$-th layer in the network structure, and $\left\{\boldsymbol{\vartheta}_{\ell, 1}, \boldsymbol{\vartheta}_{\ell, 2}\right\}$ represents the learnable parameters of the $\ell$-th layer.} \label{Fig.NearReg}
    \label{Compare_scenarios}
\end{figure*}

\subsection{Insertion Losses Optimization Based on NAG}
Due to the simplicity and efficiency in high signal-to-noise ratio (SNR) regimes, we use the zero-forcing (ZF) algorithm for DMA beamforming design. Therefore, the beamforming matrix that satisfies the maximum supplied power $P_{s}^{\text{max}}$ constraint is
\begin{equation}
\mathbf{F}_{\text{dma}}=\xi_{\text{dma}} \mathbf{H}_{\text{dma}}^{\dagger},
\end{equation}
where
\begin{equation} 
\xi_{\text{dma}}=\frac{\sqrt{P_{s}^{\text{max} }} }{\sqrt{\operatorname{Tr}\left\{\operatorname{Re}\left\{\frac{\sigma_{x}^{2}}{2}\left(\mathbf{H}_{\text{dma}}^{\dagger}\right)^{H} \mathbf{Y}_{\text{q}}\mathbf{H}_{\text{dma}}^{\dagger}\right\}\right\} }} 
\end{equation}
represents the power control factor and
\begin{equation} 
\mathbf{H}_{\text{dma}}^{\dagger}=\left(\mathbf{H}_{\text{dma}}\right)^{H}\left(\mathbf{H}_{\text{dma}}\left(\mathbf{H}_{\text{dma}}\right)^{H}\right)^{-1}
\end{equation}
corresponds to the pseudo-inverse of the DMA equivalent channel as defined in (19).
By substituting (28) into (9), we can obtain SINR of the $k$-th user as
\begin{equation}
\gamma_{k}=\gamma=\frac{P_{t}^{\max } \sigma_{x}^{2}}{\sigma_{n}^{2} \operatorname{Tr}\left\{\operatorname{Re}\left\{\frac{\sigma_{x}^{2}}{2}\left(\mathbf{H}_{\text{dma}}^{\dagger}\right)^{H} \mathbf{Z}_{\text{q}} \mathbf{H}_{\text{dma}}^{\dagger}\right\}\right\}}. \quad \forall k.
\end{equation}
Therefore, the optimization problem of DMA beamforming can be modeled as
\begin{equation}
\begin{aligned}
(\mathcal{P} 1): \quad &\underset{\mathbf{Z}_{\text{s}}}{\text{maximize}}\quad \gamma \\
&\text { subject to } \operatorname{Re}\left\{\left(\mathbf{Z}_{\text{s}}\right)_{i, i}\right\}=R_{s} \quad \forall i,
\end{aligned}
\end{equation}
where $\mathbf{Z}_{\text{s}}$ is a diagonal matrix containing the load admittance of each DMA element, with its real part denoted as $R_{s}$, indicating the presence of losses.
Notice that the objective function of $(\mathcal{P} 1)$ is differentiable, and the constraints of $(\mathcal{P} 1)$ are simple; we apply a gradient descent approach to solve the problem $(\mathcal{P} 1)$. Specifically, we employ the auto-differentiation mechanism of PyTorch to compute the derivative of the objective function with respect to the optimization variable. Furthermore, we utilize the Nesterov accelerated gradient (NAG) descent algorithm to solve it. The insertion losses optimization based on the NAG approach is summarized in Algorithm 1.

\begin{algorithm}[!t] 
	\caption{Insertion Losses Optimization Based on NAG in DMA Systems} 
	\label{alg:2} 
	\begin{algorithmic}[1] 
		\REQUIRE ~ 
	The momentum parameter $\theta$, the learning rate $\alpha$ and the number of the algorithm iteration $J$.
		\ENSURE ~ 
       DMA element load admittance matrix $\mathbf{Z}_{s}$.
       \STATE Initialize $\boldsymbol{\tau  }$ and $\gamma$.
        \FOR{$\jmath =1 \rightarrow J$}	
        \STATE Compute $\mathbf{\Delta }_{\jmath }=\theta \mathbf{\Delta}_{\jmath -1}+\alpha \nabla \gamma \left(\boldsymbol{\tau  }_{\jmath-1 }-\theta \mathbf{\Delta}_{\jmath-1 }\right)$.
      
        \STATE Update $\boldsymbol{\tau  }_{\jmath }=\boldsymbol{\tau }_{\jmath -1}-\mathbf{\Delta }_{\jmath}$
        \STATE Obtain $\mathbf{Z}_{\text{s}}=\operatorname{diag}\left(\boldsymbol{\tau  }_{1}, \ldots, \boldsymbol{\tau  }_{N_{\alpha }\times N_{\beta  }}\right)$ and update $\gamma$.
        \ENDFOR
		\RETURN $\mathbf{Z}_{\text{s}}$.
	\end{algorithmic}
\end{algorithm}
\subsection{Hybrid Beamforming Based on Unfolding PGD-Net}
In this subsection, our objective is to design hybrid precoders that maximize the SE. The optimization problem can be formulated as
\begin{subequations}
\begin{align}
(\mathcal{P} 2): \quad &\underset{\mathbf{F}_{\text{A}},\mathbf{F}_{\text{D}}}{\text {minimize}} \left\|\mathbf{F}_{\text{opt}}-\mathbf{F}_{\text{A}} \mathbf{F}_{\text{D}}\right\|_{\mathcal{F}}^{2}\\
&\text { subject to } \mathbf{F}_{\text{A}} \in \mathcal{A} \text {, }\\
&\frac{\sigma_{x}^{2}}{2} \operatorname{Tr}\left\{\operatorname{Re}\left\{(\mathbf{F}_{\text{A}}\mathbf{F}_{\text{D}}  )^{H} \mathbf{Z}_{\text{tt}} (\mathbf{F}_{\text{A}} \mathbf{F}_{\text{D}} )\right\}\right\}\le P_{t},
\end{align}
\end{subequations}
where $\mathbf{F}_{\text{opt}}$ is the unconstrained optimal digital precoder which can be obtained through the zero-forcing algorithm, i.e.,
\begin{equation}
\mathbf{F}_{\text{opt}}=\xi_{\text{fd}} \mathbf{H}_{\text{fd}}^{\dagger}
\end{equation}
with
\begin{equation} 
\xi_{\text{fd}}=\frac{\sqrt{P_{s}^{\text{max} }} }{\sqrt{\operatorname{Tr}\left\{\operatorname{Re}\left\{\frac{\sigma_{x}^{2}}{2}\left(\mathbf{H}_{\text{fd}}^{\dagger}\right)^{H} \mathbf{Z}_{\text{tt}}\mathbf{H}_{\text{fd}}^{\dagger}\right\}\right\} }}. 
\end{equation}
The analog precoder $\mathbf{F}_{\text{A}}$ is constrained as
\begin{equation}
\mathbf{F}_{\text{A}} \in \mathcal{A} \triangleq\left\{\mathbf{F}_{\text{A}} :\left[\mathbf{F}_{\text{A}} \right]_{a, b}=e^{j \varrho _{a, b}}, \forall a, b\right\},
\end{equation}
where $\varrho _{a, b}$ represents the impact of the phase shifter between the $b$-th RF chain and the $a$-th antenna.

\emph{1) Main Idea:} Based on \cite{7397861}, we propose an iterative alternating minimization approach. Specifically, in each iteration, we first neglect the constraint (31c) and then optimize $\mathbf{F}_{\text{A}}$ with $\mathbf{F}_{\text{D}}$ given. Then we design $\mathbf{F}_{\text{D}}$ to satisfy the constraint considering the fixed $\mathbf{F}_{\text{A}}$. Therefore, the first subproblem is written as
\begin{subequations}
\begin{align}
(\mathcal{P} 3): \quad &\underset{ \mathbf{F}_{\text{A}}}{\text {minimize}} \left\|\mathbf{F}_{\text{opt}}-\mathbf{F}_{\text{A}}\mathbf{F}_{\text{D}} \right\|_{\mathcal{F}}^{2}\\
&\text { subject to } \mathbf{F}_{\text{A}} \in \mathcal{A}.
\end{align}
\end{subequations}
Then, we introduce the quadratic form of the objective function without impacting the solution. Defining
\begin{subequations}
\begin{align}
&\tilde{\mathbf{n}} \triangleq \operatorname{vec}\left(\mathbf{F}_{\text{A}}\right) \in \mathbb{C}^{M N_{\text{RF}} \times 1},\\ 
&\tilde{\mathbf{m}} \triangleq \operatorname{vec}\left(\mathbf{F}_{\text{opt}}\right) \in \mathbb{C}^{M K \times 1},\\
&\tilde{\mathbf{D}} \triangleq\left(\mathbf{F}_{\text{D}}\right)^{T} \otimes \mathbf{I}_{M} \in \mathbb{C}^{M K \times M N_{\text{RF}}},
\end{align}
\end{subequations}
the re-expressed objective function is achieved and is given by
\begin{equation}
\left\|\mathbf{F}_{\text {opt }}-\mathbf{F}_{\text{A}} \mathbf{F}_{\text{D}}\right\|_{\mathcal{F}}^{2}=\|\tilde{\mathbf{m}}-\tilde{\mathbf{D}} \tilde{\mathbf{n}}\|^{2}.
\end{equation}
Furthermore, by denoting
\begin{equation}
\mathbf{n} \triangleq\left[\begin{array}{c}
\mathfrak{R}(\tilde{\mathbf{n}}) \\
\mathfrak{I}(\tilde{\mathbf{n}})
\end{array}\right] \in \mathbb{R}^{2 M N_{\text{RF}} \times 1}
\end{equation}
\begin{equation}
\mathbf{m} \triangleq\left[\begin{array}{c}
\mathfrak{R}(\tilde{\mathbf{m}}) \\
\mathfrak{I}(\tilde{\mathbf{m}})
\end{array}\right] \in \mathbb{R}^{2 MK \times 1}
\end{equation}
\begin{algorithm}[!t] 
	\caption{Offline Unsupervised Training in PGD-Net} 
	\label{alg:2} 
	\begin{algorithmic}[1] 
		\REQUIRE ~ 
	Training set $\mathcal{C}$ of channels.
		\ENSURE ~ 
       Deep unfolding network parameters $\left\{\boldsymbol{ \vartheta }_{\ell, 1}, \boldsymbol{ \vartheta }_{\ell, 2}\right\}_{\ell=1}^{L}$.
       \STATE Initialize weights $\left\{\boldsymbol{ \vartheta }_{\ell, 1}, \boldsymbol{ \vartheta }_{\ell, 2}\right\}_{\ell=1}^{L}$ and learning rate.
        \FOR{$\zeta=1 \rightarrow \mathcal{E}$}	
        \STATE Randomly divide $\mathcal{C}$ into $\mathcal{T}$ batches $\left\{\mathcal{H}^{(t)}\right\}_{t=1}^{\mathcal{T}}$.
        \FOR{$t=1 \rightarrow \mathcal{T}$}	
        \STATE Obtain $\mathbf{F}_{\text{opt}}^{(t)}$, initialize $\mathbf{F}_{\text{A}}^{(t,0)}$ with random values and obtain $\mathbf{F}_{\text{D}}^{(t,0)}$.
        \FOR{$i=1 \rightarrow \mathcal{I}_{\text {net }}^{\text {train }}$}	
        \STATE Compute $\mathbf{n}^{(t,i)}$, $\mathbf{m}^{(t,i)}$ and $\mathbf{D}^{(t,i)}$ from $\mathbf{F}_{\text{opt}}^{(t)}$.
        \STATE Compute $\overline{\mathbf{m}}^{(t, i)}=\left(\mathbf{D}^{(t, i)}\right)^{T} \mathbf{m}^{(t, i)}$.
        \STATE Compute $\overline{\mathbf{D}}^{(t, i)} \triangleq\left(\mathbf{D}^{(t, i)}\right)^{T} \mathbf{D}^{(t, i)}$.
        \STATE $\mathcal{L}^{(t, i)}=0$, $\mathbf{n}_{0}^{(t, i)}=\mathbf{0}$.
        \FOR{$\ell=1 \rightarrow L$}	
        \STATE $\mathbf{x}_{\ell-1}^{(t, i)}=-\overline{\mathbf{m}}^{(t, i)}+\overline{\mathbf{D}}^{(t, i)} \mathbf{n}_{\ell-1}^{(t, i)}$
		\STATE $\hat{\mathbf{n}}_{\ell}^{(t, i)}=\boldsymbol{ \vartheta }_{\ell, 1}^{(t, i)} \odot \mathbf{n}_{\ell-1}^{(t, i)}+\boldsymbol{ \vartheta }_{\ell, 2}^{(t, i)} \odot \mathbf{x}_{\ell-1}^{(t, i)}$
        \STATE $\mathbf{n}_{\ell}^{(t, i)}=\psi\left(\hat{\mathbf{n}}_{\ell}^{(t, i)}\right)$
		\STATE Accumulate the average loss value of the batch over PGD-Net's layers based on $\mathcal{L}^{(t, i)}=\mathcal{L}^{(t, i)}+\log (\ell) \frac{1}{ \mid \mathcal{H}^{(t) \mid}} \left\|\mathbf{m}^{(t, i)}-\mathbf{D}^{(t, i)} \mathbf{n}_{\ell}^{(t, i)}\right\|^{2}$.
        \ENDFOR
        \STATE $\mathcal{L}\left(\left\{\boldsymbol{ \vartheta }_{\ell, 1}^{(t, i)}, \boldsymbol{ \vartheta }_{\ell, 2}^{(t, i)}\right\}_{\ell=1}^{L}\right)=\mathcal{L}^{(t, i)}$.
        \STATE Update $\left\{\boldsymbol{ \vartheta }_{\ell, 1}^{(t, i+1)}, \boldsymbol{ \vartheta }_{\ell, 2}^{(t, i+1)}\right\}$ using an optimizer.
        \STATE Compute $\mathbf{F}_{\text{A}}^{(t, i)}=\mathcal{V}^{-1}\left(\mathbf{n}_{\ell}^{(t, i)}\right)$ and obtain $\mathbf{F}_{\text{D}}^{(t, i)}$.
		\ENDFOR
        \ENDFOR
        \ENDFOR
		\RETURN $\left\{\boldsymbol{ \vartheta }_{\ell, 1}, \boldsymbol{ \vartheta }_{\ell, 2}\right\}=\left\{\boldsymbol{ \vartheta }_{\ell, 1}^{\left(\mathcal{T}, \mathcal{I}_{\text {net }}^{\text {riain }}\right)}, \boldsymbol{ \vartheta }_{\ell, 2}^{\left(\mathcal{T}, \mathcal{I}_{\text {net }}^{\text {train }}\right)}\right\}$.
	\end{algorithmic}
\end{algorithm}
\begin{equation}
\mathbf{D}\triangleq\left[\begin{array}{ll}
\mathfrak{R}(\tilde{\mathbf{D}}) & -\mathfrak{I}(\tilde{\mathbf{D}}) \\
\mathfrak{I}(\tilde{\mathbf{D}}) & \mathfrak{R}(\tilde{\mathbf{D}})
\end{array}\right] \in \mathbb{R}^{2 M K\times 2 M N_{\text{RF}}},
\end{equation}
where $\mathfrak{R}(\cdot)$ and $\mathfrak{I}(\cdot)$ represent the real and imaginary parts of a complex vector or matrix, respectively, we can express
\begin{equation}
\left\|\mathbf{F}_{\text{opt }}-\mathbf{F}_{\text{A}}\mathbf{F}_{\text{D}}\right\|_{\mathcal{F}}^{2}=\|\mathbf{m}- \mathbf{D} \mathbf{n}\|^{2}.
\end{equation}
We define the transformation $\mathcal{V}: \mathbf{F}_{\text{A}} \rightarrow \mathbf{n}$ and $\mathcal{V}^{-1}: \mathbf{n} \rightarrow \mathbf{F}_{\text{A}}$.
With the introduction of the newly defined variables, the optimal solution to problem (41) can be expressed in the form of least squares (LS), i.e.,
\begin{equation}
\mathbf{n}^{\star}=\underset{\mathbf{n}: \mathcal{V}^{-1}(\mathbf{n}) \in \mathcal{A}}{\operatorname{argmin}} \|\mathbf{m}-\mathbf{D}\mathbf{n}\|^{2}.
\end{equation}
Based on (42), we can directly calculate the gradient of $\|\mathbf{m}-\mathbf{D n}\|^{2}$ with respect to $\mathbf{n}$. Therefore, a deep unfolding DNN with $L$ layers is designed to emulate the PGD algorithm and approximate $\mathbf{n}^{\star}$. As shown in equation (43), $\mathbf{n}_{\ell}$ can be generated as
\begin{equation}
\begin{aligned}
\mathbf{n}_{\ell} & =\mathcal{F}_{\ell}\left(\mathbf{n}-\mu_{\ell} \frac{\partial \|\mathbf{m}-\mathbf{D} \mathbf{n}\|^{2}}{\partial \mathbf{n}}\right)_{\mathbf{n}=\mathbf{n}_{\ell-1}} \\
& =\mathcal{F}_{\ell}\left(\mathbf{n}_{\ell-1}-\left(\mu_{\ell} \mathbf{D}^{T} \mathbf{m}+\mu_{\ell} \mathbf{D}^{T} \mathbf{D} \mathbf{n}_{\ell-1}\right)\right) \\
& =\mathcal{F}_{\ell}\left(\mathbf{n}_{\ell-1}-\mu_{\ell} \overline{\mathbf{m}}+\mu_{\ell}  \overline{\mathbf{D}} \mathbf{n}_{\ell-1}\right),
\end{aligned}
\end{equation}
where $\mu_{\ell}$ represents the learning rate and $\mathcal{F}_{\ell}(\cdot)$ denotes a nonlinear projection operator in the $\ell$-th layer. And we denote $\overline{\mathbf{D}} \triangleq \mathbf{D}^{T} \mathbf{D}$ and $\overline{\mathbf{m}} \triangleq \mathbf{D}^{T} \mathbf{m}$.
The nonlinear projection is achieved by utilizing trainable parameters, specifically the weights of the DNN. By applying this projection across multiple layers, the DNN can be designed and trained in a manner that the final output, denoted as $\mathbf{n}_{\ell}$, becomes a reliable approximation of $\mathbf{n}^{\star}$. In the subsequent sections, we introduce an efficient DNN architecture known as PGD-Net, which embodies these principles.

\begin{algorithm}[!t] 
	\caption{Online Perform in PGD-Net} 
	\label{alg:2} 
	\begin{algorithmic}[1] 
		\REQUIRE ~ 
	PGD-Net's trained parameters $\left\{\boldsymbol{\vartheta}_{\ell, 1}, \boldsymbol{\vartheta}_{\ell, 2}\right\}_{\ell=1}^{L}$.
		\ENSURE ~ 
       $\mathbf{F}_{\text{A}}$ and $\mathbf{F}_{\text{D}}$.
       \STATE Initialize $\mathbf{F}_{\text{A}}^{(0)}$ and $\mathbf{F}_{\text{D}}^{(0)}$.
        \FOR{$i=1 \rightarrow \mathcal{I}_{\text {net }}$}	
        \STATE Obtain $\mathbf{n}^{(i)}$, $\mathbf{m}^{(i)}$ and $\mathbf{D}^{(i)}$ from $\mathbf{F}_{\text{A}}^{(i-1)}$ and $\mathbf{F}_{\text{D}}^{(i-1)}$.
        \STATE Compute $\overline{\mathbf{m}}^{(i)}=\left(\mathbf{D}^{(i)}\right)^{T} \mathbf{m}^{(i)}$.
        \STATE Compute $\overline{\mathbf{D}}^{(i)} \triangleq\left(\mathbf{D}^{(i)}\right)^{T} \mathbf{D}^{(i)}$.
        \FOR{$\ell=1 \rightarrow L$}	
        \STATE $\mathbf{x}_{\ell-1}^{(i)}=-\overline{\mathbf{m}}^{(i)}+\overline{\mathbf{D}}^{(i)} \mathbf{n}_{\ell-1}^{(i)}$
		\STATE $\hat{\mathbf{n}}_{\ell}^{(i)}=\boldsymbol{ \vartheta }_{\ell, 1}^{(i)} \odot \mathbf{n}_{\ell-1}^{(i)}+\boldsymbol{ \vartheta }_{\ell, 2}^{(i)} \odot \mathbf{x}_{\ell-1}^{(i)}$
        \STATE $\mathbf{n}_{\ell}^{(i)}=\psi\left(\hat{\mathbf{n}}_{\ell}^{(i)}\right)$
        \ENDFOR
        \STATE Reconstruct the complex RF beamforming matrix $\mathbf{F}_{\text{A}}$ from $\mathbf{n}_{L}^{(i)}$ 
        \STATE When $i=1, \ldots, \mathcal{I}_{\text {net }}-1$, compute $\mathbf{F}_{\text{D}}^{(i)}$ based on (48). When $i=\mathcal{I}_{\text {net }}$, set $\mathbf{F}_{\text{A}}=\mathbf{F}_{\text{A}}^{\left(\mathcal{I}_{\text {net }}\right)}$ and obtain $\mathbf{F}_{\text{D}}$ based on (49), (50).
		\ENDFOR
	
	\end{algorithmic}
\end{algorithm}
\emph{2) PGD-Net Architecture:}
According to (43), we define the intermediate variable $\mathbf{x}_{\ell-1}$ as
\begin{equation}
\mathbf{x}_{\ell-1} \triangleq-\overline{\mathbf{m}}+ \overline{\mathbf{D}}\mathbf{n}_{\ell-1}
\end{equation}
and rewrite (44) as
\begin{equation}
\mathbf{n}_{\ell}=\mathcal{F}_{\ell}\left(\mathbf{n}_{\ell-1}+\mu_{\ell} \mathbf{x}_{\ell-1}\right)
\end{equation}
Moreover, the activation function is defined as
\begin{equation}
\psi(t,x)=-1+\frac{1}{|t|}({\sf ReLu}(x+t)-{\sf ReLu}(x-t)),
\end{equation}
where ${\sf ReLu}(\cdot)$ represents the rectified linear unit (ReLu) activation function, and $t$ denotes a hyperparameter. This ensures that the amplitudes of the elements of $\mathbf{n}_{\ell}$ fall within the range of $[-1,1]$. 
Through simulations, we have discovered that by carefully fine-tuning the hyperparameter $t$, $\psi(t, x)$ yields superior performance compared to $\tanh(x)$.

\emph{3) Training PGD-Net:} We define the loss function of the PGD-Net network as
\begin{equation}
\mathcal{L}\left(\left\{\boldsymbol{ \vartheta }_{\ell, 1}, \boldsymbol{ \vartheta }_{\ell, 2}\right\}_{\ell=1}^{L}\right)=\sum_{\ell=1}^{L} \log (\ell)\left(\left\|\mathbf{m}-\mathbf{D} \mathbf{n}_{\ell}\right\|^{2}\right)
\end{equation}
which calculates the total loss by summing the weighted objective values of all $L$ layers.
 
In Algorithm 2, we summarize the PGD-Net training process using a training data set $\mathcal{C}$.
The PGD-Net is trained over $\mathcal{E}$ epochs, each using $\mathcal{T}$ batches $\left\{\mathcal{H}^{(t)}\right\}_{t=1}^{\mathcal{T}}$.
And we have $\mathcal{H}^{(t)}=\left\{\mathbf{H}_{1}, \ldots,\mathbf{H}_{\left|\mathcal{H}^{(t)}\right|}\right\}$ where $\left|\mathcal{H}^{(t)}\right|$
denotes the training batch size. For the $t$-th batch, $\mathbf{F}_{\text{A}}^{(t,0)}$ and
\begin{equation}
\mathbf{F}_{\text{D}}^{(t,i)}=\left(\mathbf{F}_{\text{A}}^{(t,i)}\right)^{\dagger} \mathbf{F}_{\text{opt}}^{(t)}, \forall  t,i
\end{equation}
are generated randomly by using the LS method.
$\mathbf{F}_{\text{opt}}$ is the optimal fully digital precoder by ZF algorithm for the channels in $\mathcal{H}^{(t)}$. 
We begin by iteratively optimizing the PGD-Net weights from step 6.
Specifically, in the $i$-th iteration, we obtain real numbers $\mathbf{n}^{(t, i)}$, $\mathbf{m}^{(t, i)}$ and $\mathbf{D}^{(t, i)}$ by transforming complex values $\mathbf{F}_{\text{A}}^{(t, i)}$ and $\mathbf{F}_{\text{D}}^{(t, i)}$.
Steps 11 to 16 involve updating variable $\mathbf{n}_{\ell}^{(t, i)}$ and the loss value. Subsequently, in step 18, these updated values are utilized by an optimizer to update the weights $\left\{\boldsymbol{\vartheta}_{\ell, 1}^{(t, i)}, \boldsymbol{\vartheta}_{\ell, 2}^{(t, i)}\right\}_{\ell=1}^{L}$.
During the training process, each data batch is trained in an iterative process with $\mathcal{I}_{\text {net }}^{\text{train }}$ iterations. The values of variable $\mathbf{F}_{\text{A}}^{(t, i)}$ and variable $\mathbf{F}_{\text{D}}^{(t, i)}$ are constantly updated with each iteration.
This iterative approach based on deep unfolding demonstrates efficiency in reducing the training data requirement and expediting the convergence process.

\emph{4) Perform PGD-Net online:}
After the offline training process is completed, PGD-Net equipped with the trained weights is readily deployed for online beamforming design.
We summarize this process in Algorithm 3. More specifically, our first step is to generate the analog precoder and compute the digital one. The unfolding hybrid beamforming design is conducted through a series of iterations.
In steps 3-5, $\mathbf{n}^{(i)}$, $\mathbf{m}^{(i)}$ and $\mathbf{D}^{(i)}$ are obtained to compute $\overline{\mathbf{m}}^{(i)}$ and $\overline{\mathbf{D}}^{(i)}$, respectively.
After that, PGD-Net iteratively executes steps 6-10 to construct the outputs of its layers.
In step 11, the final output of PGD-Net, denoted as $\mathbf{n}_{L}$, is reconstructed as the feasible solution to $\mathbf{F}_{\text{A}}$. Additionally, the coefficients $\mathbf{F}_{\text{D}}$ are updated using the least squares (LS) method.
\begin{equation}
\mathbf{F}_{\text{D}}^{(i)}=\left(\mathbf{F}_{\text{A}}^{(i)}\right)^{\dagger} \mathbf{F}_{\text{opt}}, \forall  i
\end{equation}
Finally, with $\mathbf{F}_{\text{A}}$ obtained, the digital beamforming matrix satisfies the following power constraints, i.e.,
\begin{equation}
\mathbf{F}_{\text{D}}=\frac{\sqrt{P_{t}^{\max }} \mathbf{F}_{\text{A}}\mathbf{F}_{\text{D}}}{\sqrt{\operatorname{Tr}\left\{\operatorname{Re}\left\{\frac{\sigma_{x}^{2}}{2}\left(\mathbf{F}_{\text{A}}\mathbf{F}_{\text{D}}\right)^{H} \mathbf{Z}_{\text{tt}}(\mathbf{F}_{\text{A}}\mathbf{F}_{\text{D}})\right\}\right\}}}.
\end{equation}

\section{Power Control Design in Multi-Port Matching Networks}
In this section, we consider the power allocation problem for multi-port matching network. We focus on uplink communication, with similar extensions applicable to downlink communication.
The received signal at the BS is given by
\begin{equation}
\mathbf{y}=\mathbf{h}_{k} \sqrt{\rho_{k}} s_{k}+\sum_{i=1, i \neq k}^{K} \mathbf{h}_{i} \sqrt{\rho_{i}} s_{i}+\mathbf{n},
\end{equation}
where $\rho_{k}$ and $s_{k}$ are the transmit power and information-bearing signal of user $k$, respectively, and $\mathbf{n}$
is the additive white Gaussian noise (AWGN). With the linear receive beamforming $\mathbf{g}_{k} \in \mathbb{C}^{M \times 1}$ applied for user $k$, the received SINR is
\begin{equation}
\gamma_{k}=\frac{\rho_{k}\left|\mathbf{g}_{k}^{H} \mathbf{h}_{k}\right|^{2}}{\sum_{i=1, i \neq k}^{K} \rho_{i}\left|\mathbf{g}_{k}^{H} \mathbf{h}_{i}\right|^{2}+\frac{\sigma_{n}^{2}}{\sigma_{x}^{2}}}, \forall k.
\end{equation}
Then, the power allocation problem can be written as 
\begin{subequations}
\begin{align}
(\mathcal{P} 4): \quad &\underset{\left\{\rho_{k},\mathbf{g}_{k}: \forall k\right\}}{\text{maximize}}\quad\sum_{k=1}^{K} \log _{2}\left(1+\gamma _{k}\right)\\
&\text { subject to }\rho_{k} \leq p_{\max }, \quad k=1, \ldots, K,\\
&\frac{\sigma_{x}^{2}}{2} \operatorname{Tr}\left\{\operatorname{Re}\left\{\mathbf{g}_{k}^{H} \mathbf{Z}_{\text{rr}} \mathbf{g}_{k}\right\}\right\}=1, \quad k=1, \ldots, K.
\end{align}
\end{subequations}

Specifically, constraint (53b) is the maximum transmit power constraint, where $p_{\max }$ denotes the maximum transmit power available for each user. 
Moreover, (53c) ensures that the beamforming matrix satisfies regularization constraints.
Generally, this problem can be solved using WMMSE \cite{5756489}. Nevertheless, this kind of approach exhibits high computational complexity. To this end, we use the deep unfolding network to unroll the conventional WMMSE algorithm by introducing learnable parameters. 
Furthermore, we employ GNN as generic layers, which serve as a natural extension of CNN to the graph domain, enabling effective processing of non-Euclidean data. Unlike multi-layer perceptron (MLP) and CNNs, GNN can be directly extended to various network scenarios without the need for retraining, which is a significant advantage \cite{10296858}. 

We define the effective channel as $\bar{h}_{k,m}=\mathbf{h}_{k}^{H} \mathbf{f}_{m}$.
\begin{figure}[t]
	\centering
	\includegraphics[width=0.4\textwidth]{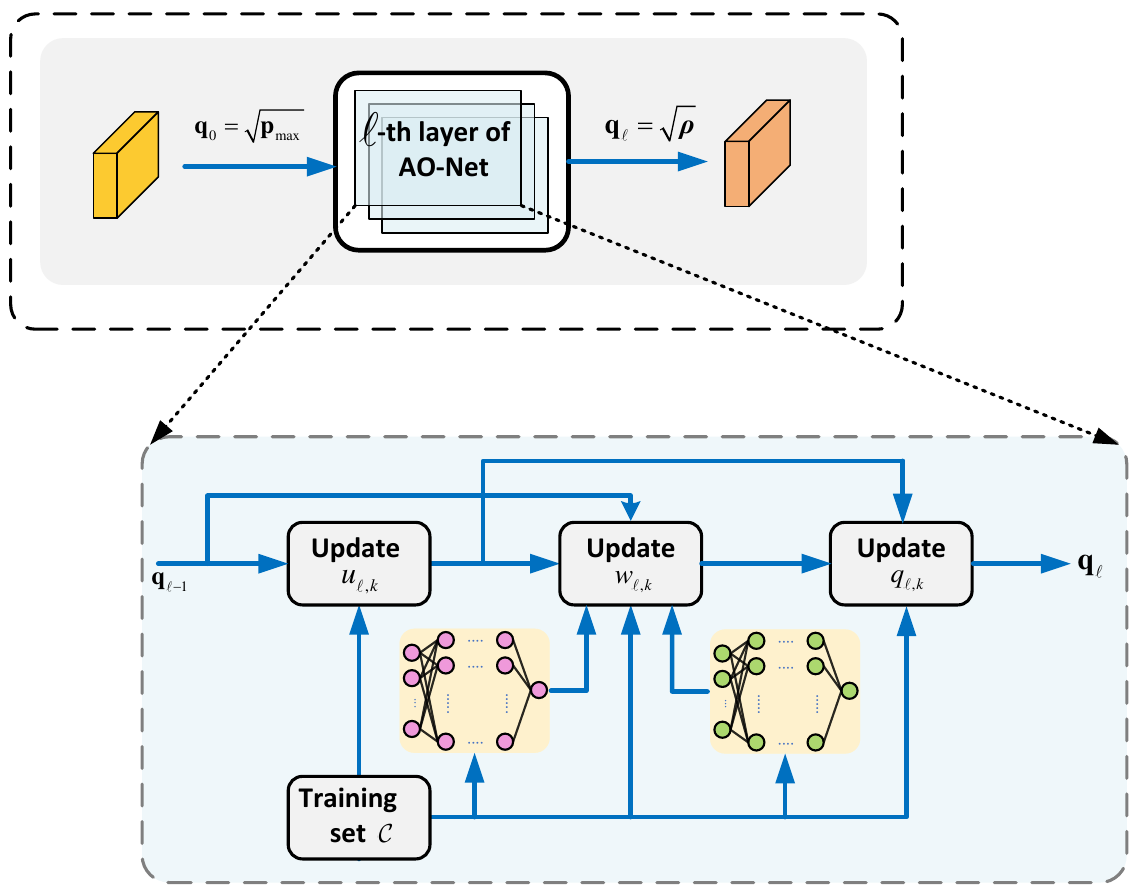}
	\caption{The architecture of the proposed GNN aided AO-Net for power allocation design. 
The input to the layered structure is $\mathbf{q}_{0}=\sqrt{\mathbf{p}_{\text {max }}}$ and the computed power allocation is given by $\mathbf{q}_{\ell}=\sqrt{\boldsymbol{\rho}}$.
The parameters $\boldsymbol{\vartheta}_{\ell, 1}$ and $\boldsymbol{\vartheta}_{\ell, 2}$ for all layers $\ell$ are learned to minimize the loss $\mathcal{L}\left(\left\{\boldsymbol{\vartheta}_{\ell, 1}, \boldsymbol{\vartheta}_{\ell, 2}\right\}_{\ell=1}^{L}\right)$, thus promoting a faster convergence than its classical WMMSE counterpart.} \label{Fig.NearReg}
    \label{AO_Net}
\end{figure}
As shown in Fig. \ref{AO_Net}, the relationship for the $\ell $-th layer of the unfolded network can be expressed as follows
\begin{equation}
\mathbf{a}_{\ell }=\boldsymbol{\Theta} \left(\mathbf{H} ; \boldsymbol{\vartheta }_{\ell ,1}\right), \quad \mathbf{b}_{\ell }=\boldsymbol{\Theta} \left(\mathbf{H} ; \boldsymbol{\vartheta }_{\ell ,2}\right),
\end{equation}
\begin{equation}
u_{\ell ,k}=\frac{\bar{h}_{k,k} q_{\ell-1,k}}{\frac{\sigma_{n}^{2}}{\sigma_{x}^{2}}+\sum_{m} \bar{h}_{k,m}^{2} q_{\ell-1,m}^2}, \quad k=1, \ldots, K,
\end{equation}
\begin{equation}
w_{\ell,k}=\frac{a_{\ell ,k}}{1-u_{\ell,k} \bar{h}_{k,k} q_{\ell ,k}^{(k-1)}}+b_{\ell ,k}, \quad k=1, \ldots, K,
\end{equation}
\begin{equation}
q_{\ell ,k}=\nu \left(\frac{u_{\ell ,k} \bar{h}_{k,k} w_{\ell ,k}}{\sum_{m} \bar{h}_{m,k}^{2} u_{\ell,m} u_{\ell ,m} w_{\ell ,m}}\right), \quad k=1, \ldots, K,
\end{equation}
where the nonlinear function $\nu (x ):=[x]_{0}^{\sqrt{p_{\max }}}$ in equation (58) is employed to constrain $q_{\ell ,k}$ within the interval $\left[0, \sqrt{p_{\max }}\right]$, effectively saturating the right-hand side of equation (8) at these boundary values.
$\boldsymbol{\Theta}(\cdot )$ denotes the GNN architecture.
The trainable parameters $\boldsymbol{\Psi}$ is given by $\boldsymbol{\Psi}=\left[\boldsymbol{\vartheta }_{1 ,1}, \boldsymbol{\vartheta }_{1 ,2}, \ldots, \boldsymbol{\vartheta }_{L ,1}, \boldsymbol{\vartheta }_{L ,2}\right]$.
Furthermore, we define the loss function, denoted as
\begin{equation}
\mathcal{L}\left(\left\{\boldsymbol{\vartheta}_{\ell, 1}, \boldsymbol{\vartheta}_{\ell, 2}\right\}_{\ell=1}^{L}\right)=-\sum_{\ell=1}^{L}\log (\ell) \sum_{k=1}^{K} \log _{2}\left(1+\gamma_{k,\ell }\right).
\end{equation}
We summarize the pseudo-code of the GNN-aided AO-Net in Algorithm~4.
\begin{algorithm}[!t] 
	\caption{Unsupervised Training in GNN-Aided AO-Net} 
	\label{alg:2} 
	\begin{algorithmic}[1] 
		\REQUIRE ~ 
	Training set $\mathcal{C}$ of channels.
		\ENSURE ~ 
The parameters of the DNN Network $\left\{\boldsymbol{ \vartheta }_{\ell, 1}, \boldsymbol{ \vartheta }_{\ell, 2}\right\}_{\ell=1}^{L}$.
       \STATE Initialize DNN weights $\left\{\boldsymbol{ \vartheta }_{\ell, 1}, \boldsymbol{ \vartheta }_{\ell, 2}\right\}_{\ell=1}^{L}$ and learning rate.
        \FOR{$\zeta =1 \rightarrow \mathcal{E}$}	
        \STATE Randomly divide $\mathcal{C}$ into $\mathcal{T}$ batches $\left\{\mathcal{H}^{(t)}\right\}_{t=1}^{\mathcal{T}}$.
        \FOR{$b=1 \rightarrow \mathcal{B}$}	
        \FOR{$\ell=1 \rightarrow L$}
        \STATE Compute $\mathbf{a}_{\ell }=\boldsymbol{\Theta} \left(\mathbf{H} ; \boldsymbol{\vartheta }_{\ell ,1}\right)$ appling the GNN weights	$\boldsymbol{\vartheta }_{\ell ,1}$.
        \STATE Compute $\mathbf{b}_{\ell }=\boldsymbol{\Theta} \left(\mathbf{H} ; \boldsymbol{\vartheta }_{\ell ,2}\right)$ appling the GNN weights	$\boldsymbol{\vartheta }_{\ell ,2}$.
        \STATE Compute $u_{\ell,k}$ according to (56).
		\STATE Update $w_{\ell,k}$ according to (57).
        \STATE Obtain $q_{\ell, k}$ according to (58).
		\STATE Accumulate the average loss value of the batch over AO-Net's layers based on $\mathcal{L}^{(t)}=\mathcal{L}^{(t)}-\log (\ell) \frac{1}{ \mid \mathcal{H}^{(t) }\mid}\sum_{k=1}^{K} \log _{2}\left(1+\gamma_{k,\ell }\right)$.
        \ENDFOR
        \STATE $\mathcal{L}\left(\left\{\boldsymbol{ \vartheta }_{\ell, 1}^{(t)}, \boldsymbol{ \vartheta }_{\ell, 2}^{(t)}\right\}_{\ell=1}^{L}\right)=\mathcal{L}^{(t)}$.
        \STATE Obtain $\left\{\boldsymbol{ \vartheta }_{\ell, 1}^{(t)}, \boldsymbol{ \vartheta }_{\ell, 2}^{(t)}\right\}$ with an optimizer.
        \STATE Update $\mathbf{a}_{\ell }$, $\mathbf{b}_{\ell }$, $u_{\ell,k}$, $w_{\ell,k}$ and $q_{\ell, k}$.
		\ENDFOR
        \ENDFOR
		\RETURN $\left\{\boldsymbol{ \vartheta }_{\ell, 1}, \boldsymbol{ \vartheta }_{\ell, 2}\right\}=\left\{\boldsymbol{ \vartheta }_{\ell, 1}^{\left(\mathcal{T}\right)}, \boldsymbol{ \vartheta }_{\ell, 2}^{\left(\mathcal{T}\right)}\right\}$.
	\end{algorithmic}
\end{algorithm}
\section{Simulation Results}
In this section, we explore the impact of electromagnetic phenomena, e.g., insertion losses, mutual coupling, and reflections inside the waveguides on the DMA performance of multi-port matching networks. Moreover, we evaluate the performance of the proposed beamforming and power allocation scheme through numerical simulations.

\subsection{Simulation Setup}
We consider a single-cell communication scenario in which one BS equipped with FD arrays, hybrid digital/analog arrays, or DMAs serves multiple single-antenna UEs.
We set the frequency $f$ to $10$ Hz. 
The dimensions of the DMA waveguides are characterized by width $a=0.73 \lambda$ and height $b = 0.167 \lambda$, respectively, where $\lambda$ represents the wavelength. 
The simulation parameters are disposed in TABLE \ref{table}.

\begin{table}[t]
\caption{The Simulation Parameters in Multi-Port Matching Network.}
\vspace{4mm}
\centering
\fontsize{9}{10}\selectfont
\begin{tabular}{|c|c|}
    \hline
    \hline
    \bf Parameters & \bf Values \cr\hline
    Number of waveguides/RF chains $N_{\text{RF}}$ & 6  \cr\hline
    Number of antennas in each waveguide $N_{\mu}$ & 20  \cr\hline
    Number of antennas $M$ & 120  \cr\hline
    Frequency $f$& 10 GHz  \cr\hline
    Number of UEs $K$ & 6  \cr\hline
    Power consumption of a single RF chain $P_{\text{RF}}$ & 250 mW \cite{10161727} \cr\hline
    Power consumption of a single phase shifter $P_{\text{PS}}$& 10 mW  \cr\hline
    Connector intrinsic impedance $Z_{0}$& 35.3387 S  \cr\hline
    The parasitic resistance $R_{s}$ & 0 \cr\hline
    The width of the waveguide $a$ & 0.73$\lambda $ \cr\hline
    The height of the waveguide $b$ & 0.167$\lambda $ \cr\hline
    The length of the waveguide $L_{\mu}$ & 110 mm \cr\hline
    Transmitted power $P_{t}$ $/$ supplied power $P_{s}$ & 1 W \cr\hline
    Number of layers $L$ & 4 \cr\hline
    Iterations of training $\mathcal{I}_{\text {net }}^{\text {train }}$& 1 \cr\hline
    \hline
\end{tabular}
\label{table}
\end{table}
\subsection{Performance Analysis of Multi-port Matching Network}
\begin{figure}[t]
	\centering
	\includegraphics[width=0.35\textwidth]{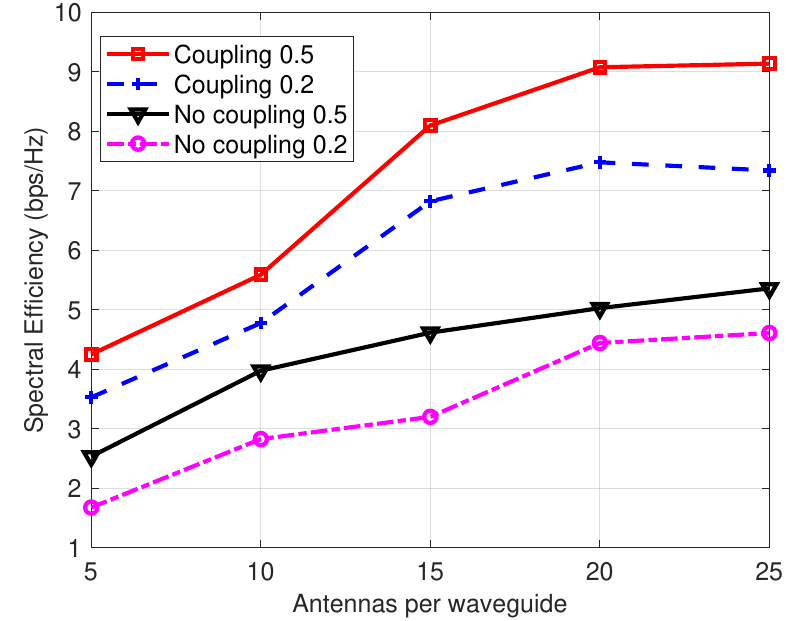}
	\caption{Impact of mutual coupling in the spectral efficiency for the DMA system. $K=6$ users are served by $N=6$ RF chains with different element numbers per waveguide.} \label{Fig.NearReg}
    \label{sim_2}
\end{figure}

As shown in Fig. \ref{sim_2}, we investigate the impact of mutual coupling on the system performance, with a focus on two distinct inter-element spacing parameters: $0.5\lambda$ and $0.2\lambda $. 
It is evident that a complete neglect of the coupling phenomenon leads to a substantially degraded performance.
Mutual coupling affects the radiation patterns of antennas and the spatial correlation of channels. Therefore, when designing the system's precoding matrix, ignoring the mutual coupling effect will cause antenna mismatch and reduce the system's SE compared to considering it.
Meanwhile, we observe that the DMA system performs better with half wavelength antenna spacing.
These illustrate that the antenna parameters of radio channel models have substantial impacts on channel capacity. Therefore, it is necessary to consider the
antenna parameters in channel models.
Another intriguing observation emerges: there is an upper limit on the attainable rate within the DMA system.
This observation aligns with the fundamental principles governing the functionality of a DMA. The power injected into the waveguides is partially radiated by the initial element, as dictated by the termination admittance. The remaining power continues to propagate until it reaches the subsequent element, where the process repeats. It is intuitively evident that, with an increasing number of elements, progressively diminishing amounts of power reach the final elements. Consequently, this implies the existence of an optimal number of elements per waveguide that strikes a balance between cost and performance.

\begin{figure}[t]
	\centering
	\includegraphics[width=0.35\textwidth]{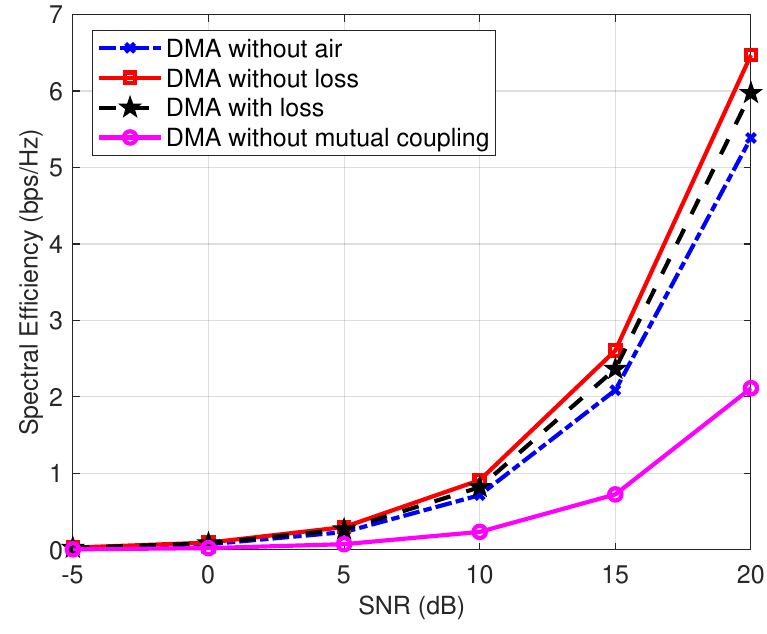}
	\caption{Impact of environmental parameters on performance. $K=6$ users are served by $N=6$ RF chains of the DMA system with different SNRs.} \label{Fig.NearReg}
    \label{sim_1}
\end{figure}

In Fig. \ref{sim_1}, we examine the influence of insertion losses on the DMA system. The following are the differences between several curves:

$\bullet$ \emph{DMA without air:} The coupling between the elements through the air is neglected, i.e., the value of $G_{e 2, z z}^{(a)}$ is neglected.

$\bullet$ \emph{DMA with loss:} Based on the algorithm 1 in Section III, we can obtain the result of the DMA with loss.

$\bullet$ \emph{DMA without loss:} Setting $\boldsymbol{\Lambda}=\mathbf{0}$ in equation (20) yields the performance obtained by the same algorithm without considering the reflection coefficient during optimization.

$\bullet$ \emph{DMA without mutual coupling:} The consideration of mutual coupling is entirely disregarded, meaning that $\mathbf{Z}_{\text{ss}}$ is constrained to be a diagonal matrix.

DMA optimization performance with insertion loss (black curve) is close to that ignoring insertion loss (red curve), which proves the feasibility of our NAG algorithm.
Based on the results, it is evident that a complete disregard for coupling can result in notably inadequate performance. Conversely, disregarding coupling through air can be considered a reasonable approximation only under the circumstance where the spacing between components is substantial and does not exert a substantial influence on performance.
The conclusions are well-founded and illustrate the importance of modeling the mutual coupling both through waves and air in DMA-compatible systems.


\subsection{Comparison of Beamforming Design}
The objective of this subsection is to compare the SE and EE performance between the DMA-based and conventional antenna-assisted systems, considering both FD and hybrid architectures. To quantify the EE, we adopt the energy consumption model and define the parameter EE as
\begin{equation}
\text{EE}=\frac{\text{SE}}{P_{\text{total }}},
\end{equation}
where $P_{\text {total }}$ represents the total power consumption of the BS. The total power consumption of the FD architecture is given by
\begin{equation}
P_{\text {total }}^{\text{FD}}=P_{t}+MP_\text{RF},
\end{equation}
where $P_{t}$ is the transmitted power and $P_\text{RF}$ represents the power consumption of single RF chain. The power consumption of the fully-connected architecture is
\begin{equation}
P_{\text {total }}^{\text{FC}}=P_{t}+N_\text{RF} P_\text{RF}+N_\text{RF} M P_\text{PS},
\end{equation}
where $P_\text{PS}$ is the single phase shifter's power consumption. The power consumption of the sub-connected architecture is defined as
\begin{equation}
P_{\text {total }}^{\text{SC}}=P_{t}+N_{\text{RF}} P_{\text{RF}}+MP_{\text{PS}}.
\end{equation}
The total energy consumed by the DMA structure is
\begin{equation}
P_{\text {total }}^{\text{DMA}}=P_{s}+N_\text{RF} P_\text{RF},
\end{equation}
where $P_{s}$ is the supplied power of the DMA system.
\begin{figure}[t]
	\centering
	\includegraphics[width=0.35\textwidth]{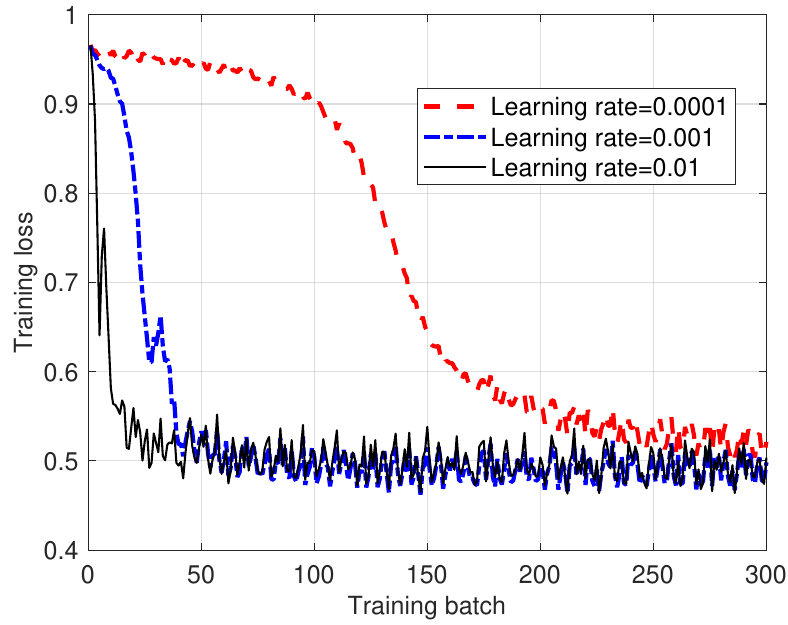}
	\caption{Normalized training loss of PGD-Net with different learning rates for multi-port matching networks  with $N_{\mu}=20$, $N_{\text{RF}}=6$ and $K=6$.} \label{train}
    \label{sim_4}
\end{figure}
\begin{figure}[t]
	\centering
	\includegraphics[width=0.35\textwidth]{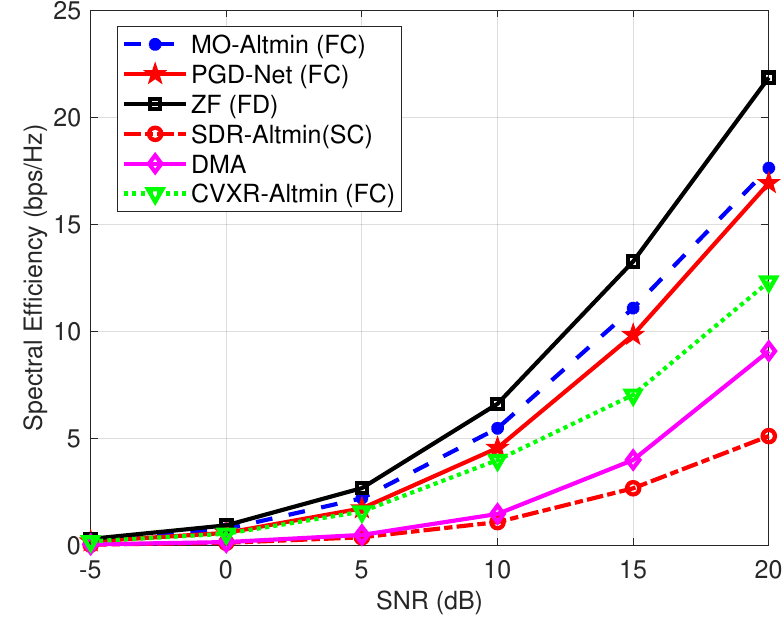}
	\caption{Spectral efficiency comparison of the proposed PGD-Net beamforming algorithm for multi-port matching networks with $N_{\mu}=20$, $N_{\text{RF}}=6$ and $K=6$.} \label{Fig.NearReg}
    \label{sim_3_1}
\end{figure}
\begin{figure}[t]
	\centering
	\includegraphics[width=0.35\textwidth]{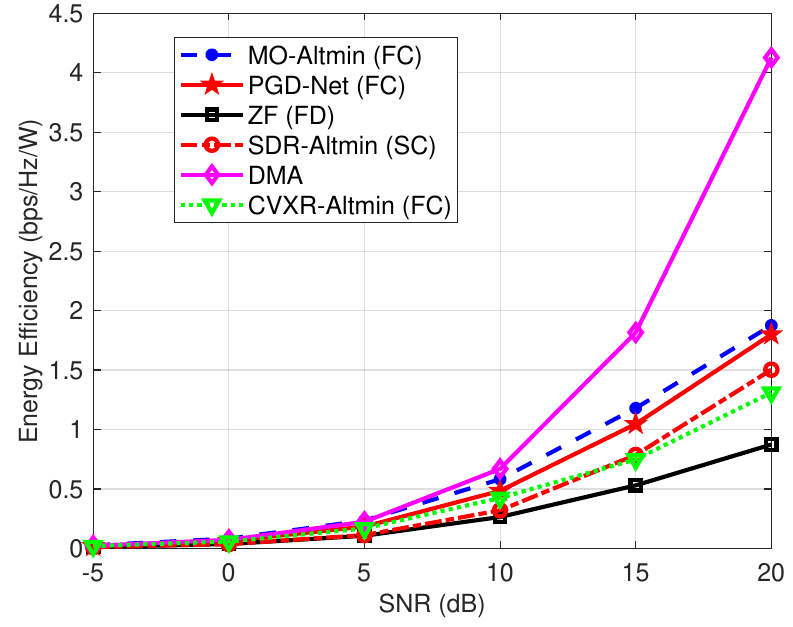}
	\caption{Energy efficiency comparison of the proposed PGD-Net beamforming algorithm for multi-port matching networks with $N_{\mu}=20$, $N_{\text{RF}}=6$ and $K=6$.} \label{Fig.NearReg}
    \label{sim_3_2}
\end{figure}
\begin{figure}[t]
	\centering
	\includegraphics[width=0.35\textwidth]{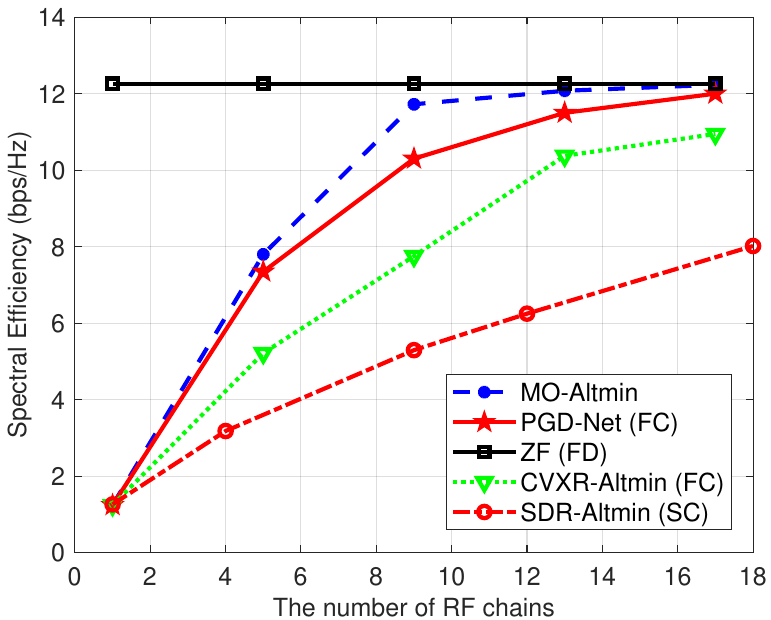}
	\caption{Spectral efficiency performance versus the number of RF chains for multi-port matching networks with $M=120$, $K=6$ and $\text{SNR}=20$ dB.} \label{SE_RF}
    \label{sim_5}
\end{figure}

Fig. \ref{train} demonstrates the average training loss versus batches under different learning rates. For investigated learning rates, it is observed that the loss decreases at different rates and eventually converges. Therefore, we choose $\mu_{\ell}=0.01$ as the simulation parameter.

Fig. \ref{sim_3_1} and Fig. \ref{sim_3_2} present the SE and EE performances under different SNRs for multi-port matching networks, respectively.
To showcase the performance achieved by our proposed algorithms, we compare them against some baseline algorithms. The specific details are outlined below.

$\bullet$ \emph{ZF (FD):}  ZF algorithm is designed for the fully-digital architecture as the performance upper bound.

$\bullet$ \emph{MO-Altmin (FC) \cite{[68]}:} MO-Altmin optimization is designed for the fully-connected architecture to address non-convex constraints in constant mode in problem (6).

$\bullet$ \emph{CVXR-Altmin (FC) \cite{7579557}:} Similar to MO-Altmin algorithm, CVXR-Altmin is designed for the fully-connected architecture with lower complexity.

$\bullet$ \emph{SDR-Altmin (SC) \cite{7397861}:} SDR-Altmin optimization is designed for the sub-connected architecture to address non-convex constraints in constant mode in problem (6).

It is evident that digital beamforming exhibits higher SE performance compared to hybrid beamforming operations.
The main idea of the PGD-Net is to unfold the MO-AltMin algorithm. By comparing the results of the PGD-Net method to those of the MO-Altmin algorithm, we can effectively gauge the effectiveness and efficiency of the proposed approach.
Although MO-Altmin optimization and PGD-Net algorithm both achieve almost an identical SE, our algorithm has a lower complexity, as will be further investigated in next subsection. 
Based on the observations depicted in Fig. \ref{sim_3_2}, it is evident that the DMA-assisted architecture outperforms the conventional fully digital architecture in terms of EE due to the reduced number of RF chains. Furthermore, the EE performance of the DMA-assisted architecture surpasses that of the fully-connected hybrid A/D one. This discrepancy arises from the fact that the hybrid A/D architecture necessitates additional power to accommodate numerous phase shifters, whereas DMAs do not require any supplementary circuitry to execute signal processing in the analog domain. As anticipated, the EE performance of the hybrid A/D architecture surpasses that of the fully digital architecture, which aligns with the reduced number of RF chains inherent to the hybrid A/D architecture.

When the number of data streams (i.e., $N_{\text{RF}}$) is fixed, the
number of RF chains can be varied in the investigated algorithms. In Fig. \ref{SE_RF}, we study the impact of the number of RF
chains in multi-port matching networks, where $M = 120$, $K=6$ and $\text{SNR}=20$ dB, respectively. We observe that when the number of RF chains is 16, its performance approximates the fully-digital beamforming structure. 
In addition, it performs better than CVXR-Altmin algorithm and sub-connected structures based on SDR-Altmin algorithm.

\begin{figure}[t]
	\centering
	\includegraphics[width=0.3\textwidth]{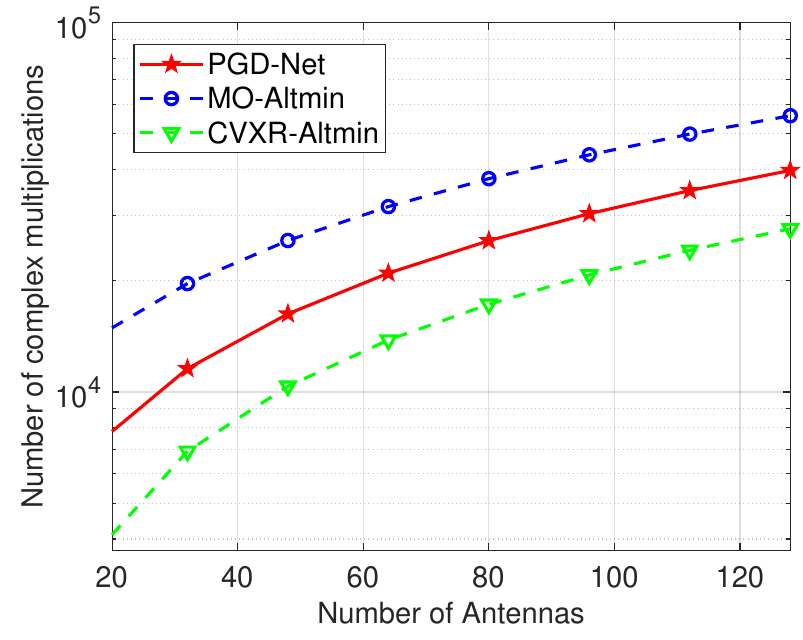}
	\caption{Number of complex multiplications versus the number of antennas.} \label{Complexity_1}
    \label{sim_5}
\end{figure}
\begin{figure}[t]
	\centering
	\includegraphics[width=0.3\textwidth]{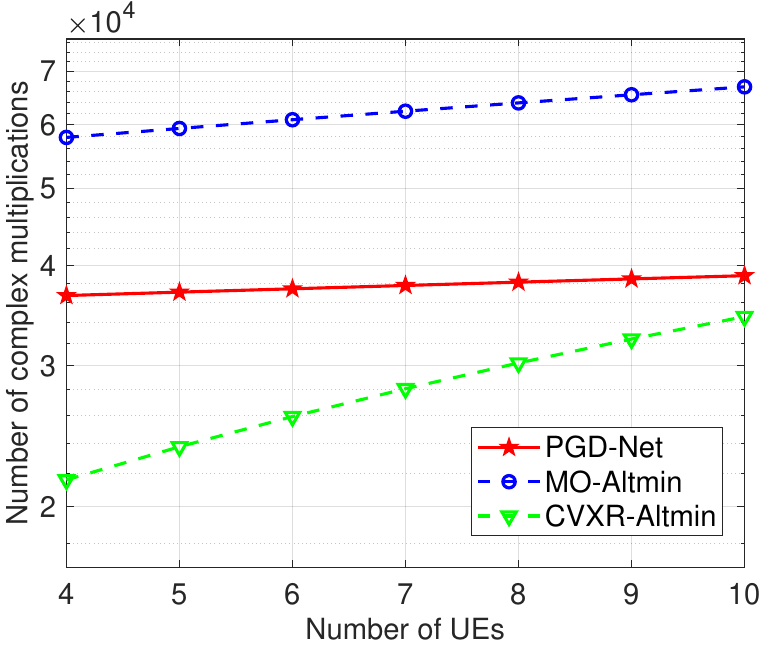}
	\caption{Number of complex multiplications as a function of the number of UEs.} \label{Complexity_2}
    \label{sim_5}
\end{figure}

\subsection{Comparison of Complexity}

In Fig. \ref{Complexity_1} and \ref{Complexity_2}, we present a complexity analysis of the investigated algorithms, including MO-Altmin algorithms and our proposed PGD-Net in multi-port matching networks.
The computational complexity required by the PGD-Net algorithm is about 29.68\% times lower than traditional MO-Altmin algorithm.
While the PGD-Net exhibits higher complexity in comparison to the traditional CVXR-Altmin algorithm \cite{7579557}, it is evident that as the number of users increases, the complexity of the CVXR-Altmin algorithm experiences a significant surge. This observation signifies that our algorithm possesses enhanced scalability. Moreover, our algorithm demonstrates superior performance.

\subsection{Comparison of Power Allocation Design}
In this subsection, we consider the power allocation problem in multi-port matching network with $M=36$ and $K=30$.
We mainly compare the proposed GNN aided AO-Net with the following three benchmarks:

$\bullet$ \emph{WMMSE \cite{5756489}:} For $K$-users interference channel power control, this is the most popular optimization algorithm.

$\bullet$ \emph{CNN \cite{8444648}:} To learn near-optimal power control, CNN and the unsupervised loss function are used.

$\bullet$ \emph{Equal power:} We assume that each user is assigned equal power.

As show in TABLE \ref{power}, it is evident that conventional CNN networks are unsuitable for optimizing multi-port impedance matching network systems. In contrast, our proposed AO-Net can strike a balance between complexity and performance, approaching the traditional WMMSE scheme. It is not enough to achieve a near-optimal sum-rate since fading conditions have to be synchronized with the time requirement for power allocation.
As a result, it is pertinent to take into account the duration required by each algorithm to generate a power allocation output given a channel state input. We also compare the computation times of the various algorithms in TABLE \ref{power}. AO-Net, with a processing time of approximately $10$ ms per sample, exhibits a significant speed advantage compared to WMMSE, which requires approximately $61$ ms per sample.
Furthermore, we investigate the impact of the number of layers on system performance. Increasing the number of layers in a network has the potential to enhance performance. Nevertheless, it concurrently amplifies network complexity and runtime duration.
\begin{table}[t]
 \caption{Average Spectral Efficiency and Running Time for Investigated Algorithms.}\label{power}
 \centering
 \footnotesize
 \ra{1.5}
\begin{tabular}{ l c c c }
  \toprule
   &   {Performance (bps/Hz)} &   {Test time (ms)} \\
  \midrule
  
 WMMSE & 33.2980 & 61.0 \\
 \rowcolor{lightblue}
  AO-Net (4 layers) & 32.7180  &10.0 \\
  \rowcolor{lightblue}
AO-Net (6 layers)  & 33.1270 &28.0  \\
CNN & 25.0906 & 13.5  \\
  Equal power & 31.6776  & -  \\
  \bottomrule
\end{tabular}
\end{table}
\section{Conclusion}
This paper presents a performance analysis of multi-port matching network systems. According to the results, DMAs are more energy efficient than mMIMO systems. In addition, the DMA optimization must account for insertion losses, with particular attention to the significant impact of mutual coupling, primarily within the waveguides. Next, we introduce a deep unfolding framework, i.e., PGD-Net, which leverages unfolding projected gradient descent for the hybrid beamforming design of the multi-port matching network. Furthermore, we develop an innovative deep unfolding network, GNN aided AO-Net, to learn how to rapidly and robustly optimize power, while being fully interpretable. Numerical results demonstrate that the proposed PGD-Net based hybrid beamforming approaches approximate the conventional model-based algorithm with very low complexity. In the future, we will study more efficient beamforming algorithms in DMA systems.

%
%
%
%
%
%

\begin{appendices}
\section{Computational Complexity Analysis of the Investigated Methods}\label{secA}
We delve into the computational complexity of the proposed PGD-Net scheme outlined in Algorithm 3.
It is observed that matrix $\mathbf{B}$ exhibits sparsity, where each row and column contain only $2N_{\text{RF}}$ and $2K$ nonzero real-valued elements, respectively.
The computational complexity for calculating $\overline{\mathbf{m}}$ and $\overline{\mathbf{D}}$ is only $\mathcal{O}\left(K N_{\text{RF}}\right)$ and $\mathcal{O}\left( N_{\text{RF}}^{2} K\right)$, respectively.
Furthermore, $\overline{\mathbf{D}}$ has only $2N_{\text{RF}}$ nonzero elements in each row and column, and hence step 7 requires a complexity of $\mathcal{O}\left(M+2 K N_{\text{RF}}\right)$.
The weighting operation in step 8 involves only element-wise vector multiplication or addition, resulting in a complexity of $3 \mathcal{O}\left(M N_{\text{RF}}\right)$.
In step 12, obtaining $\mathbf{F}_{\text{D}}^{(i)}$ with (49) has a complexity of $\mathcal{O}\left(M N_{\text{RF}}^{2}\right)$.
Consequently, the total complexity of Algorithm 3 can be estimated as
\begin{equation}
\begin{array}{l}
\mathcal{C}_{\text {PGD-Net }}=\left(\mathcal{I}_{\text {net }}-1\right) \mathcal{O}\left(M  N_{\text{RF}}^{2}\right)+\mathcal{O}\left(M  N_{\text{RF}}\right) \\
\quad+\mathcal{I}_{\text {net }} \mathcal{O}\left(2  N_{\text{RF}}^{2} K+L\left(3 M N_{\text{RF}}+2  N_{\text{RF}} K\right)\right) .
\end{array}
\end{equation}

Compared to MO-Altmin, the proposed PGD-Net based beamforming scheme has low complexity. The MO-Altmin algorithm requires complexities of
\begin{equation}
\begin{array}{l}
\mathcal{C}_{\text {MO-AltMin }}= \\
\quad \mathcal{I}_{\mathrm{MO}}^{\text {out }} \mathcal{O}\left(MN_{\text{RF}}^{2}+\mathcal{I}_{\text{MO}}^{\text {in }}\left(3 M N_{\text{RF}}+2 \left(N_{\text{RF}}^{2}+N_{\text{RF}}\right) K\right)\right).
\end{array}
\end{equation}
SDR-AltMin requires complexities of $\mathcal{O}\left(M K\right)$ and $\mathcal{O}\left( K^{3} N_{\text{RF}}^{3}\right)$ to calculate the analog and digital precoders of each iteration. In total, the complexity of the SDR-Altmin is
\begin{equation}
\mathcal{C}_{\text {SDR-AltMin }}=\mathcal{I}_{\text {SDR }} \mathcal{O}\left(M K+K^{3} N_{\text{RF}}^{3}\right).
\end{equation}
\end{appendices}

\bibliographystyle{IEEEtran}
\bibliography{IEEEabrv,Ref}

\end{document}